\documentclass[twocolumn,10pt,a4paper,float,
amsmath, amssymb, aps,
pra
]{revtex4-1}

\usepackage[utf8]{inputenc}

\usepackage[dvipsnames]{xcolor}
\usepackage{hyperref}
\hypersetup{
    colorlinks=true,
    citecolor=BrickRed,
    linkcolor=RoyalBlue
}

\usepackage{float}
\usepackage{graphicx}% Include figure files
\usepackage{dcolumn}% Align table columns on decimal point
\usepackage{bm}
\usepackage{mathrsfs}

\newcommand{\la}{\langle}
\newcommand{\ra}{\rangle}
\newcommand{\nex}{n_{\text{ex}}}
\newcommand{\Nex}{N_{\text{ex}}}

\begin{document}

\title{\large Bose-Einstein condensate fluctuations versus an interparticle interaction}

\author{S.V. Tarasov$^1$, Vl.V. Kocharovsky$^1$, and V.V. Kocharovsky$^{1,2}$\\
\textit{$^{1}$Institute of Applied Physics, Russian Academy of Science, Nizhny Novgorod 603950, Russia}\\
\textit{$^{2}$Department of Physics and Astronomy, Texas A\&M University, College Station, TX 77843-4242, USA}
}
\date{\today}

\begin{abstract}
We calculate the Bose-Einstein condensate (BEC) occupation statistics vs. the interparticle interaction in a dilute gas with a nonuniform condensate in a box trap within the Bogoliubov approach. The results are compared against the previously found BEC-occupation statistics in (i) an ideal gas and (ii) a weakly interacting gas with a uniform condensate. In particular, we reveal and explicitly describe an appearance of a nontrivial transition from the ideal gas to the Thomas-Fermi regime. The results include finding the main regimes of the BEC statistics -- the anomalous non-Gaussian thermally dominated fluctuations and the Gaussian quantum dominated fluctuations -- as well as a crossover between them and their manifestations in a mesoscopic system. Remarkably, we show that the effect of the boundary conditions, imposed at the box trap, on the BEC fluctuations doesn't vanish in the thermodynamic limit of a macroscopic system even in the presence of the interparticle interactions. Finally, we discuss a challenging problem of an experimental verification of the theory of the BEC fluctuations addressing a much deeper level of the many-body statistical physics than usually studied quantities related to the mean condensate occupation.   

% PACS numbers: 03.75.Hh, 64.60.an, 05.70.Fh, 05.70.Ln
\end{abstract}

\maketitle

%%%%%
%%%%%
%%%%%

\section{A challenge of predicting and measuring the BEC fluctuations} 
\label{sec: intro}

A significant progress in the experimental studies of Bose-Einstein condensation in dilute gases \cite{Kristensen2019,MehboudiPRL2019,Hadzibabic2017QD,ChangPRL2016,Dalibard2015,Perrin2012,toroidBEC-PRL2011,Hadzibabic2010,Armijo2010,Jacqmin2010,Campbell,Hung2011,Cornell2010,Esslinger2007} during a quarter of century since its experimental realization
\cite{Cornell1995,Ketterle1995,Ketterle1999,Dalibard2008} opens a possibility to directly measure Bose-Einstein condensate (BEC) fluctuations. 
Understanding and testing such fluctuations mean reaching a much deeper level of quantum statistical physics of phase transitions and critical phenomena than a level of mean BEC and quasiparticle characteristics adopted in the bulk of studies (see, e.g., \cite{Hadzibabic2017QD,ChangPRL2016,PitString2016,Steinhauer2002,Makotyn2014,Pieczarka2020}). 
The point is that an order parameter in the continuous phase transitions arises at a critical temperature via very large, critical fluctuations (e.g., see \cite{Kuklov2006}).
Besides, knowing and controlling the BEC fluctuations are important for various BEC applications. For instance, in the matter-wave interferometers (like Ramsey \cite{DrummondPRA2019,Drummond2011} or Mach-Zehnder \cite{Chip1000atoms} on-chip interferometers), which complement optical interferometers in precision measurement devices, the BEC occupation fluctuations directly affect the phase diffusion arising from the interactions between the atoms. 
There are many deep physical problems relevant to the atom-number fluctuations in the BEC-condensed systems, including the atom-number difference between two halves of a trapped Bose gas \cite{Castin} which is generic for the matter-wave interferometers, the number fluctuations in small cells of quantum gases determining deviations from the thermodynamic limit fixed by the isothermal compressibility \cite{Pit2011}, etc. In the present paper we consider only fluctuations in the total occupation of the condensate or noncondensate.

Recently, a breakthrough progress in a direct observation of the BEC-occupation fluctuations, including the sudden increase in BEC fluctuations close to the critical temperature, has been achieved in \cite{Kristensen2019} (for other experiments on the BEC fluctuations, see \cite{Armijo2010,Jacqmin2010,Hung2011,Cornell2010,Esslinger2007,RaizenBECstatisticsPRL2005,AspectDensityFluctPRL2006,SchmittPhotonBECStatisticsPRL2014,StoofBECofLightPRL2014}). The advance in \cite{Kristensen2019} is based on a stabilization technique, which allows for the preparation of ultracold thermal clouds at the shot noise level and elimination of numerous technical noise sources, as well as making use of the correlations established by the evaporative cooling process to precisely determine the fluctuations and the sample temperature. 

A direct control and variation of the interparticle interactions in the experiments could be done via a well established technique based on tuning the Feshbach resonances \cite{Campbell,Chin2010,Kohler2006}. 

A full theoretical description of the BEC occupation statistics, that gives the moments higher than the mean value and is valid for all temperatures including an entire critical region near a critical temperature $T_c$ of the BEC phase transition, is known only for an ideal gas \cite{PRA2010,PRA2014,JStatPhys2015}. 
Finding such a theory for an interacting gas remains a difficult unsolved problem, especially in the critical region of parameters. 
The first investigations of the BEC fluctuations in a weakly interacting gas could be traced back well before the modern era of the dilute-gas BEC in the magneto-optical traps (in particular, see \cite{Pule1983}).  
Some results are available for a weakly interacting gas, but only at relatively low temperatures and within a mean-field Bogoliubov approach. 
In particular, an increase of the BEC fluctuations due to switching from a standard thermodynamic, Gaussian scaling in an ideal gas to an anomalous scaling in a weakly interacting gas had been predicted in \cite{Pit1998} and studied in \cite{LiuBECfluctHarmonicTrapPRA2003} for a nonuniform condensate in a harmonic trap.
An opposite effect of a decrease of the BEC fluctuations from an already anomalous level in an ideal gas to a two-fold lower level in a weakly interacting gas due to squeezing of fluctuations in each pair of counterpropagating coupled modes of excitations had been found in \cite{PRA2000} for a uniform condensate in a box trap with the periodic boundary conditions. 

Recently, a correlated potential harmonics expansion (CPHE) approximation \cite{Bhattacharyya2016-1} for the energy levels of a mesoscopic Bose system, that is, a perturbative generalization of a mean-field theory based on an inclusion of the two-particle correlations evaluated on a mean-field background of all other particles being the spectators, complimented by an iteration of a well-known recursion relation for a partition function of an ideal gas with a given energy levels had been applied in \cite{Bhattacharyya2016-2} for calculating fluctuations of the condensate in the harmonic trap and provided interesting results for the parameters mimicking the JILA experiment with $^{87}{\rm Rb}$ atoms. 
Beyond mean-field effects and a sharp fall in the variance of fluctuations for a large-size condensate near the critical temperature also had been discussed in \cite{Bhattacharyya2016-2}, though an accuracy of such an approach in the critical region is not clear. 

A classical-field approximation combined with the Monte Carlo method was employed to numerically study the statistical properties of the cold interacting bosons in a quasi-one-dimensional ring and harmonic traps at finite temperatures in \cite{RzazewskiPRA2020,RzazewskiPRA2011,RzazewskiPRA2009,RzazewskiJPhysB2007}. 
It replaces creation and annihilation operators by complex c-number amplitudes and neglects modes of momenta higher than a suitable cutoff momentum. 

There is another method which is suitable for computing the effect of the interparticle interactions on the BEC fluctuations in the mesoscopic systems at finite temperatures and is based on the c-number representation of the Heisenberg quantum-mechanical, operator equations and a truncated Hilbert space \cite{Drummond2019}. 
It utilizes a Wigner or positive-P phase-space representation of quantum mechanics which has provided excellent results for computing various properties of an interacting gas under the conditions close to the actual BEC experiments (e.g., see \cite{DrummondPRA2019,Drummond2019,DrummondPhysScr2016,Drummond2013,Drummond2012}). 
It would be very interesting to see predictions of this method for the potential experiments on the equilibrium and nonequilibrium BEC statistics in an interacting gas confined in different traps. 

In fact, since the original Bogoliubov theory of BEC \cite{Bogoliubov1947,Bogoliubov1967}, first formulated for the zero temperature, there were many works suggesting various approaches towards extending the theory to the higher, finite temperatures. 
One of the most advanced approaches is based on the Matsubara (temperature) Green's functions and Dyson-Beliaev equations as is described, for example, in \cite{AGD,LL,Shi1998} and papers \cite{Shlyapnikov1998,MullerPRA2015}. 
Some other approaches are described in \cite{Zwerger2004,Andersen2004,Proukakis2008} and include also a phenomenological renormalization group approach \cite{Goldenfeld1992,Bijlsma1996,Vicari2002}. 
However, none of them gives a consistent microscopic description of a continuous transition of a real mesoscopic system through the critical point, and we won't elaborate on this topic (for its discussion and a possible microscopic theory, see \cite{PLA2015,PhysScr2015,Entropy2020}). 

Let us just point to the following two facts directly relevant to the BEC fluctuations. 
First, the BEC fluctuations are closely related to a system's specific heat which is commonly discussed in the literature in this context and, for the uniform BEC, has been calculated in \cite{Popov1965,PopovBook1983} (for superfluids, see, for example, \cite{HeFilmDirichletBC1997,NhoPRB2003,HeFilmExperiment2000}). 
Indeed, according to the macroscopic thermodynamics \cite{LLV}, the specific heat involves the second derivative of an entropy (and, hence, a partition function) with respect to the temperature that, in terms of the fluctuations, corresponds to the second moment or cumulant of the BEC-occupation statistics. 
In particular, the famous Ginzburg-Levanyuk criterion which defines the critical region of a phase transition \cite{Levanyuk1959,Ginzburg1960} has been formulated in terms of the specific heat, but actually specifies a region of the system's parameters where the fluctuations of the order parameter are critical -- larger than the mean value of the order parameter.      

The second fact is a presence of the infrared divergences in the intermediate integrals of the diagrammatic expansion for the self-energies in the thermodynamic limit \cite{Gavoret1964,Bijlsma1996,Castellani1997,Shi1998}. 
These infrared divergences cancel out in the final expressions for the mathematical expectation of thermodynamic physical quantities and the long-wavelength excitation spectrum, but their origin looks similar to the origin of the infrared divergences in a formal integral representation of a mesoscopic sum for the variance of the BEC fluctuations which is anomalously large due to a contribution of a relatively small number of the lowest energy quasiparticle excitations. 
It is a manifestation of the fact that the critical, very large fluctuations are responsible for the critical phenomena and should be explicitly analyzed in order to reach a deeper level of many-body physics. 
The relation of the anomalous BEC fluctuations to the infrared divergencies in the one-particle Green’s function is well-known in the hydrodynamic theory of Bose superfluids (see, for instance, \cite{Pit1992}). 
Below, we'll describe the anomalous BEC fluctuations in an interacting gas at finite temperature in detail.  

An approach developed in the present paper for the analysis of the fluctuations in the occupation of a non-uniform Bose condensate in a dilute weakly interacting equilibrium gas is consistent with the mean-field Popov approximation which, in addition to the Bogoliubov approximation, takes into account the interactions between the noncondensed, excited particles \cite{Popov1965,PopovBook1983,Shi1998}. 
Namely, it takes into account the first order correction to the self-energy via a perturbation theory but, contrary to the self-consistent Hartree-Fock-Bogoliubov (HFB) approximation, omits the off-diagonal self-energies. 
It is a simplified and, at the same time, improved version of the HFB approximation since it also treats the excited particles as the renormalized single quasiparticles, but does not introduce a false energy gap in the excitation spectrum. 
The latter is crucially important for the correct analysis of the BEC fluctuations presented below. 
The Popov approximation allows one to consider all finite temperatures where, as per the Ginzburg-Levanyuk criterion \cite{Levanyuk1959,Ginzburg1960}, the mean-field approximation is valid, $T_c - T \gg (n^{1/3}a) T_c$, i.e., all temperatures except the critical region in a close vicinity of the critical temperature. 
As usual, a weak interparticle interaction in a dilute gas of $N$ Bose particles confined in a box of a size $L$ is characterized by a small parameter $na^3 \ll 1$ in terms of the s-wave scattering length $a$ and the gas density $n=N/L^3$.  

Yet, for simplicity's sake, in this paper we discuss only the low temperature region, $T \ll T_c$, when the mean occupation of the excited energy levels is much less than the mean condensate occupation, $\la N_{\text{ex}} \ra \ll \la N_0 \ra$, and the effect of the interactions between the noncondensed particles can be ignored. 
In this case, the Popov approximation is reduced to just Bogoliubov approximation and we can simplify, accordingly, the Gross-Pitaevskii equation for the macroscopic wave function of the condensate and the Bogoliubov-de Gennes equations for the quasiparticle profiles (see Eq.~(\ref{GP+BdG}) below) by omitting small, relative to the condensate density $n_0$, corrections to their appropriate terms due to the noncondensate density $\nex$ and its inhomogeneity. 
In other words, in this paper we calculate accurately the condensate depletion itself, $\la \nex \ra = n - \la n_0 \ra \ll n$, but neglect the corrections due to this small depletion in the factors like $\la n_0 \ra a^3 \approx na^3$.  

An extension of this analysis to the entire temperature range of validity of the Popov approximation, $T_c - T \gg (n^{1/3}a) T_c$,  with a full account for a significant nonuniform depletion of the condensate will be discussed elsewhere. Yet, already the Bogoliubov approximation allows us to reveal the main regimes in the evolution of the BEC-fluctuations statistics with the interparticle interaction parameter $na^3$ increasing from the zero value in the ideal gas regime all the way to a relatively large values ensuring an onset of the Thomas-Fermi regime that, in the general case, corresponds to a significant restructuring of the spatial profiles of the condensate and quasiparticles. 

The major goal of this paper is to calculate, analytically and numerically, a pronounced effect of the interparticle interaction on the fluctuations of the {\it nonuniform} BEC and to show that its mechanism originates in the rectructuring of the spectrum and particle content of quasiparticles via the two competing ways: (i) directly through the Bogoliubov-type coupling and squeezing \cite{PRA2000} and (ii) indirectly through the accompanying restructuring of the condensate spatial profile. 
Contrary to the previous works, like \cite{Pit1998,LiuBECfluctHarmonicTrapPRA2003} and \cite{Entropy2018}, we don't rely on the ad hoc model assumptions, e.g., about the structure and spectrum of the quasiparticles. Instead, we consistently derive all properties, including the resulting BEC statistics, of the interacting gas with the nonuniform condensate from the first principles and, at last, prove that the anomalous scaling of and effect of the boundary conditions on the BEC-occupation fluctuations are not washed out by the interparticle interactions even in the thermodynamic limit of a macroscopically large system.  

The contents of the paper is as follows. In Sect.~II we formulate a model and the basis equations to be analyzed. 
In Sect.~III we present the formulas describing how the spectrum, profile, and particle content of quasiparticles change in the course of their restructuring due to the interparticle interactions.
In Sect.~IV we analytically calculate the characteristic function and, hence, the cumulants (moments) for the statistics of the BEC occupation by means of the Wigner function technique. 
In Sect.~V we introduce a diagonal approximation for the quasiparticles and BEC statistics which is based on a BEC-modified Schr$\ddot{\text{o}}$dinger equation and constitutes the main tool for obtaining the results and numerical graphs for the BEC fluctuations presented in the paper. 

In Sect.~VI, we apply the general theory developed in the previous sections to explicitly reveal all main regimes of the occupation statistics for the nonuniform condensate in a dilute gas with the interparticle interaction ranging from the case of a vanishing interaction in the ideal gas to the case of a quite intensive interaction in the Thomas-Fermi limit. 
In particular, we describe qualitatively and calculate numerically a regime of the thermally dominated, non-Gaussian fluctuations sensitive to the boundary conditions and controlled by a relatively small group of the low energy excitations, a regime of the quantum dominated, Gaussian fluctuations, and a crossover from the anomalous to Gaussian statistics. 
An interplay between these regimes in a mesoscopic system is also analyzed. 
The conclusions and prospects for the experimental studies of the BEC fluctuations in the nonuniform interacting gas are discussed in Sect.~VII.   

The analysis in the sections~V-VII is performed for a special kind of a trap, namely, a box with the periodic boundary conditions along two Cartesian axes and the Dirichlet (zero) boundary conditions along the third axis. (For confined superfluids, such boundary conditions had been employed for the calculation of the specific-heat scaling function in the thick helium films near the critical temperature by means of the Monte Carlo simulations \cite{HeFilmDirichletBC1997,NhoPRB2003} and nicely fitted the experimental data \cite{HeFilmExperiment2000} with no free parameter.)
In such a system the condensate and quasiparticle wave functions are nonuniform and quite dependent on the interparticle scattering, as is the case for the actual experimental setups, which is, in contrast to the model of the uniform condensate in a box with all periodic boundary conditions, usually considered in the literature. 
At the same time, the chosen model is still simple enough for describing both the quasiparticles and the statistics of the total condensate occupation in a quite transparent and almost analytical way. 
Importantly, in this way we show that the effect of the boundary conditions on the BEC fluctuations found previously in an ideal gas \cite{PRA2014,JStatPhys2015} is still present in a gas with a quite intense interparticle scattering, including the Thomas-Fermi regime, even in the thermodynamic limit.

\section{The framework: Nonuniform vs. uniform condensates in the box traps within the Bogoliubov approximation}

We consider a weakly interacting dilute Bose gas of $N$ particles confined in a 3D-box trapping potential
\begin{equation}    \label{U}
    U_{\text{tr}}({\bf r}) =  \begin{cases}
                            &0,  \quad \ {\bf r} \in [0,L] \times [0,L] \times [0,L], \\
                            &\infty, \quad {\bf r} \not\in [0,L] \times [0,L] \times [0,L].
                        \end{cases}
\end{equation}
We assume the Dirichlet (zero) boundary conditions along the axis $x$ and the periodic boundary conditions along the other two axes $y$ and $z$. 
Hence, the condensate wave function is uniform in the $y$ and $z$ directions, but significantly changes in the $x$ direction depending on the interparticle scattering amplitude -- it varies from the half-sine period in the case of a very weak interaction to a half of the square-wave period in the Thomas-Fermi limit.
At the same time, the system is close to a standard textbook case of a homogeneous gas (corresponding to all periodic boundary conditions), that makes evaluating the effects of boundary conditions more transparent.

We consider low temperatures, $T \ll T_c$, and employ the mean-field Bogoliubov approach. 
Namely, the condensate and quasiparticles' wave functions are described by the Gross-Pitaevskii and Bogoliubov-de Gennes equations, respectively, which both include the average density of the condensate.
In this case the grand-canonical and canonical ensembles for the noncondensate are fully equivalent \cite{JStatPhys2015}. 
The small corrections due to the nonuniform density profile of the noncondensate could be taken into account via the Popov approximation, but we don't discuss them here since they don't change the overall picture qualitatively.

The gas is in a thermal equilibrium described by the field operator $\hat{\psi}$, statistical operator (density matrix) $\hat{\rho}$ and Hamiltonian $\hat{H}$ of the independent quasiparticles:
\begin{equation}	\label{H}
\begin{split}
    &   \hat{\rho} = e^{- \hat{H}/ T} \prod_j(1-e^{-E_j/T}), 
        \quad 
        \hat{H} = \sum_j E_j \hat{b}_j^{\dagger} \hat{b}_j;
    \\
    &\hat{\psi}({\bf r}) = \sqrt{\la N_0 \ra} \phi({\bf r}) + \hat{\psi}_{\text{ex}}({\bf r}), \\
    &\hat{\psi}_{\text{ex}}({\bf r}) = \sum_j \left( u_j({\bf r}) \hat{b}_j + v_j^*({\bf r}) \hat{b}_j^{\dagger} \right).
\end{split}
\end{equation}
The condensate part of $\hat{\psi}$ is replaced by a $c$-number macroscopic wave function $\sqrt{\la N_0 \ra} \phi$, while the noncondensate part is described by the Bose-field operator $\hat{\psi}_{\text{ex}}$. 
The latter is written in terms of the quasiparticle operators $\hat{b}_j^\dagger$, $\hat{b}_j$ which create, annihilate the quasiparticle in an excited state $j$ of an energy $E_j$. 
The wave functions of the condensate $\phi$ (which is chosen to be real) and quasiparticles $(u_j,v_j)$ are described by the Gross-Pitaevskii and Bogoliubov-de Gennes equations, respectively:
\begin{equation}    \label{GP+BdG}
\begin{split}
    &\hat{\mathscr{L}} \phi = 0,   
    \\
    &\begin{cases}
    &\hat{\mathscr{L}} u_j + g \la N_0 \ra \phi^2({\bf r}) (u_j+v_j) = + E_j u_j,   
    \\
    &\hat{\mathscr{L}} v_j + g \la N_0 \ra \phi^2({\bf r}) (u_j+v_j) = - E_j v_j;
    \end{cases}
    \\
    &\hat{\mathscr{L}} \equiv -\frac{\hbar ^2 \Delta}{2M} + U_{\text{tr}}({\bf r}) + g \la N_0 \ra \phi^2 ({\bf r}) + 2g \nex ({\bf r}) -\mu,
\end{split}
\end{equation}
where $\Delta$ is the three-dimensional Laplace operator. 
They are normalized as follows: $\int \phi^2({\bf r}) d^3{\bf r} = 1$ and $\int \left( |u_j({\bf r})|^2-|v_j({\bf r})|^2 \right) d^3{\bf r} = 1$.
Here $g=4\pi \hbar^2 a/M$ is an interaction constant, $M$ is a particle mass, $\mu$ is a chemical potential, $n_0 ({\bf r}) \equiv  \la N_0 \ra \phi^2({\bf r})$ and $ \nex ({\bf r}) = \la \hat{\psi}^{\dagger}_{\text{ex}}({\bf r}) \hat{\psi}_{\text{ex}}({\bf r}) \ra$ are the mean density profiles of the condensed and noncondensed particle fractions;
\begin{equation}    \label{n}
\nex({\bf r}) = \sum_j \Big[ |v_j({\bf r})|^2 + \frac{|u_j({\bf r})|^2 + |v_j({\bf r})|^2}{\exp (E_j/T) -1} \Big].
\end{equation}
The angles $\la \ldots \ra = \text{Tr} ( \ldots \hat{\rho})$ mean a statistical averaging.

The quantity in question is a probability distribution $\rho_0 (N_0)$ of finding $N_0$ particles in the condensate. 
Equivalently, one can look for a probability distribution $\rho_{\text{ex}}(\Nex)$ of finding the complimentary number $N_{\text{ex}} = N - N_0$ of the noncondensed particles being in the excited energy states, since the total number of particles $N$ in the trap is fixed.
These probability distributions can be found via a characteristic function $\Theta(u)$ as follows
\begin{equation}	\label{CF_PDF} 
\begin{split}
    &\Theta(u) =    \la e^{iu\hat{N}_{\text{ex}}} \ra
                    \equiv \text{Tr} \big(e^ {iu \hat{N}_{\text{ex}}} \hat{\rho}\big),  
    \\
    &\rho_{\text{ex}}(\Nex) = \frac{1}{2 \pi} \int_{-\pi}^{\pi} e^{-iu\Nex} \Theta (u)  du, 
    \\
    &\rho_0(N_0) = \rho_{\text{ex}}(N - N_0). 
\end{split}
\end{equation}
Here $\hat{N}_{\text{ex}} = \int \hat{\psi}^{\dagger}_{\text{ex}}({\bf r}) \hat{\psi}_{\text{ex}}({\bf r}) d^3{\bf r}$ is the operator of the total number of excited particles. Calculating $\Theta (u)$ is not a trivial task.
Indeed, the operator $\exp \big(iu\hat{N}_{\text{ex}}\big)$ is diagonal in a particle basis which is not the same as a basis for quasiparticles diagonalizing the density matrix $\hat{\rho}$. 
Thus, a straightforward calculation of the trace $\text{Tr}$ in Eq.~(\ref{CF_PDF}) means dealing with a bunch of overlapping integrals involving all the $u_j$ and $v_j$ wave functions as well as with a non-commuting operators in the exponent.

The method of calculating the characteristic function for an arbitrary trap is presented in Sect.~IV below.

\section{Restructuring of quasiparticles due to the interparticle interaction: Spectrum, profile, particle content}

Let us use a freedom to choose an arbitrary complete set of basis wave functions in the decomposition of the field operator over the particle states, $\hat{\psi}_{\text{ex}}({\bf r})=\sum _{k \neq 0} f_k({\bf r})\hat{a}_k$, in order to simplify a relation between the creation/annihilation operators of the particles ($\hat{a}_k^\dagger$, $\hat{a}_k$) and quasiparticles ($\hat{b}_j^\dagger$, $\hat{b}_j$) beforehand. 
A usual, trivial choice of the bare particle basis states associated with the single-particle wave functions of an empty trap with the external potential in Eq.~(\ref{U}) (for example, see \cite{Hadzibabic2019excitations}) is not productive since it doesn't allow one to easily separate the wave functions of excitations from the condensate wave function and requires to deal with a complex orthogonalization procedure for every value of the interaction. 
Much more convenient set of the basis functions is generated by the BEC-modified single-particle Schr$\ddot{\text{o}}$dinger equation \cite{WuGriffin1996,HZF}, $\hat{\mathscr{L}}[\phi({\bf r})] f = \epsilon f$, defined by the differential operator $\hat{\mathscr{L}}$ in Eq.~(\ref{GP+BdG}) for a given condensate wave function $\phi ({\bf r})$.
This choice looks natural for the basis equation since it corresponds to the main diagonal part of the Bogoliubov-de Gennes system of equations (\ref{GP+BdG}). 

Hereinafter, for the sake of simplicity, we restrict the analysis to the low temperatures, $T \ll T_c$, and neglect the depletion corrections in the factors involving the mean condensate occupation assuming $\la N_0 \ra \simeq N$ as is stated in Sect.~I. So, we employ the BEC-modified Schr$\ddot{\text{o}}$dinger equation in the following form:
\begin{equation}    \label{mod_Shr}
    - \frac{\hbar^2}{2M}\Delta f_k+\left[ U_{\text{tr}}({\bf r})+g N \phi^2({\bf r}) -\mu \right] f_k =\epsilon_k f_k.
\end{equation}
The effective potential in Eq.~(\ref{mod_Shr}) includes an additional term $g N \phi^2({\bf r})$ which depends on the condensate profile and modifies the external potential $U_{\text{tr}}({\bf r})$.
(In principle, the small noncondensate correction $2 g \nex$  to the effective potential  may be taken into account via an iterative scheme presented in \cite{HZF}).
The ground state of Eq.~(\ref{mod_Shr}) coincides with the condensate wave function, $f_0({\bf r}) = \phi({\bf r})$, and has the zero energy, $\epsilon_0 = 0$. 
All other solutions (taken to be real-valued functions) form a complete orthonormal basis $\{f_k |{k\neq0}\}$ of a subspace of the single-particle wave functions which is orthogonal to the condensate wave function and, thus, spans the wave functions of excitations.
Introducing the creation and annihilation operators $\hat{a}_k^{\dagger}$ and $\hat{a}_k$ of these basis states $\{f_k\}$, which are already dressed by the condensate, we represent the excited part of the particle field operator as $\hat{\psi}_{\text{ex}}({\bf r})=\sum _{k \neq 0} f_k({\bf r})\hat{a}_k$. 
The operator of the total number of noncondensed particles has a usual form 
\begin{equation}   \label{Nex-aa}
\hat{N}_{\text{ex} }=\sum _{k \neq 0} \hat{a}_k^\dagger \hat{a}_k.
\end{equation}

Introducing the expansions of the quasiparticle wave functions $u_j({\bf r}) = \sum_k \mathscr{A}_{jk} f_k({\bf r})$ and $v_j({\bf r}) = \sum_k \mathscr{B}_{jk} f_k({\bf r})$ over this basis, we rewrite the Bogoliubov-de Gennes equations (\ref{GP+BdG}) in the matrix form
\begin{widetext}
\begin{equation}    \label{BdG_matrix}
\mathscr{M}
\left[
\begin{array}{c}
    \mathscr{A}_{j1} \\
    \mathscr{B}_{j1} \\
    \cdots \\
    \mathscr{A}_{jk} \\
    \mathscr{B}_{jk} \\
    \cdots
 \end{array}
\right]
\equiv
\left[
\begin{array}{cccccc}
    \epsilon_1 + \Delta_{11}& 
    \Delta_{11}& 
    \cdots&
    \Delta_{1k}& 
    \Delta_{1k}& 
    \cdots
\\
    -\Delta_{11}& 
    -\epsilon_1 - \Delta_{11}& 
    \cdots&
    -\Delta_{1k}& 
    -\Delta_{1k}& 
    \cdots
\\
    \cdots & \cdots & \cdots & \cdots & \cdots & \cdots
\\
    \Delta_{k1}& 
    \Delta_{k1}& 
    \cdots&
    \epsilon_k + \Delta_{kk}& 
    \Delta_{kk}& 
    \cdots
\\
    -\Delta_{k1}& 
    -\Delta_{k1}& 
    \cdots&
    -\Delta_{kk}& 
    -\epsilon_k - \Delta_{kk}& 
    \cdots
\\
    \cdots & \cdots & \cdots & \cdots & \cdots & \cdots 
\end{array}
\right] \left[
\begin{array}{c}
    \mathscr{A}_{j1} \\
    \mathscr{B}_{j1} \\
    \cdots \\
    \mathscr{A}_{jk} \\
    \mathscr{B}_{jk} \\
    \cdots
 \end{array}
\right] =
E_j 
\left[
\begin{array}{c}
    \mathscr{A}_{j1} \\
    \mathscr{B}_{j1} \\
    \ldots \\
    \mathscr{A}_{jk} \\
    \mathscr{B}_{jk} \\
    \ldots
 \end{array}
\right],
\end{equation}
\end{widetext}
where
\begin{equation}    \label{Dij}
\Delta_{jk} = g N \int f_j \phi^2 f_k \ d^3{\bf r}
\end{equation}
are the overlapping integrals.
Note that the projection of the Bogoliubov-de Gennes equations (\ref{GP+BdG}) on each basis function $f_k$ provides two equations into the system of the algebraic equations (\ref{BdG_matrix}). 
Thus, the Bogoliubov-de Gennes matrix $\mathscr{M}$ has an intrinsic ($2\times2$)-block structure.

Summing up the two equations in the $k$-th pair of Eqs.~(\ref{BdG_matrix}), we obtain the relation $\mathscr{A}_{jk} / \mathscr{B}_{jk} = (\epsilon_k + E_j) / (\epsilon_k - E_j)$ that looks similar to the usual Bogoliubov's one for the uniform condensate. 
(Note, however, that the quasiparticle energies $E_j$ remain unknown here.)
This fact effectively halves the dimension of the system of Eqs.~(\ref{BdG_matrix}) and reduces the problem to the following eigenvalue problem:
\begin{equation}    \label{BdG_mode_mixing}
    \left[
\begin{array}{ccc}
  \epsilon_1^2+2\Delta_{11}\epsilon_1 & 
    2\Delta _{12}\sqrt{\epsilon_1\epsilon_2} & 
    \cdots
\\
  2\Delta_{21}\sqrt{\epsilon_1\epsilon_2} & 
    \epsilon _2^2+2\Delta _{22}\epsilon _2 & 
    \cdots
\\
  \cdots & \cdots & \cdots
\end{array}
\right] \!\! \left[
\begin{array}{c}
\!    p_{j1} \\
\!    p_{j2} \\
\!    \cdots
 \end{array}
\! \right] \! = 
E_j^2 \!
\left[
\begin{array}{c}
\!    p_{j1} \\
\!    p_{j2} \\
\!    \cdots
 \end{array}
\! \right] \!\!.
\end{equation}
Then, the quasiparticle eigenfunctions can be found as
\begin{equation}	\label{QP}
    (u_j,v_j) = \sum_{k \neq 0} p_{jk} 
                \left(  \frac{\epsilon_k + E_j}{2\sqrt{\epsilon_k E_j}},  
                        \frac{\epsilon_k - E_j}{2\sqrt{\epsilon_k E_j}}
                \right) f_k({\bf r}).
    % \quad
    % \cosh \xi_{??} =  \frac{\epsilon_j + \lambda_k}{2\sqrt{\epsilon_j\lambda_k}},
    % \            
    % \sinh \xi_{??} =  \frac{\epsilon_j - \lambda_k}{2\sqrt{\epsilon_j\lambda_k}}
\end{equation}

The diagonalized Bogoliubov-de Gennes matrix, $R^{-1} \mathscr{M} R$, and the matrix $R$ performing this diagonalization have the following ($2\times2$)-block structures:
\begin{equation}    \label{R}
\begin{split}
    &R^{-1} \mathscr{M} R 
        =  \!\! \left[  \begin{matrix}
                        E_1 \sigma_z  &  0 &  \cdots  \\
                        0  &  E_2 \sigma_z &  \cdots  \\
                        \cdots  &  \cdots  &  \cdots  \\
                    \end{matrix} \right]\!\!,   
    \
    \sigma_z =  \left[  \begin{matrix}  
                            +1  &   0 \\ 
                            0   &  -1 
                        \end{matrix} \right]\!\!,
    \\[5pt]
    &R = \! \left[ \begin{matrix}
                p_{11} L_{11}   &   p_{21} L_{21}   &   \cdots  \\
                p_{12} L_{12}   &   p_{22} L_{22}   &   \cdots  \\
                \cdots  &  \cdots  &   \cdots
            \end{matrix} \right]\!\!,    
    \
    L_{jk} = \! \left[  \begin{matrix}  
                            \cosh \xi_{jk} & \sinh \xi_{jk} \\ 
                            \sinh \xi_{jk} & \cosh \xi_{jk}
                        \end{matrix} \right]\!\!.
\end{split}
\end{equation}
Here $L_{jk}$ is the hyperbolic (Lorentz) rotation matrix with the entries $\cosh \xi_{jk} = \frac{\epsilon_k+E_j}{2 \sqrt{E_j \epsilon_k}}$ and $\sinh \xi_{jk} = \frac{\epsilon_k-E_j}{2 \sqrt{E_j  \epsilon_k}}$ involving \ $\exp \xi_{jk} = \sqrt{\epsilon_k/E_j}$; the coefficients $p_{jk}$ form an orthogonal matrix.
Eq.~(\ref{BdG_matrix}) and its reduction, Eq.~(\ref{BdG_mode_mixing}), obey a symmetry of the Bogoliubov-de Gennes Eq.~(\ref{GP+BdG}). Namely, any its solution $(u_j,v_j;+E_j)$ is accompanied by a counterpart, an unphysical solution $(v_j, u_j;-E_j)$ with a negative norm and a negative energy.

The matrix $R$ constitutes the Bogoliubov transform describing a transformation from the particle operators $\hat{a}_k^{\dagger}$ and $\hat{a}_k$ to the quasiparticle operators $\hat{b}_j^{\dagger}$ and $\hat{b}_j$, as it follows from the two representations of the field operator of the excited particles:
\begin{equation}
   \hat{\psi}_{\text{ex}}({\bf r}) 
    =  \sum_{k\neq0} f_k({\bf r}) \hat{a}_k =
   \sum_j \Big( u_j({\bf r}) \hat{b}_j + v_j^*({\bf r}) \hat{b}_j^{\dagger} \Big).
\end{equation}
In the general case, finding the Bogoliubov transform implies finding two pairs of the complex conjugated functions, $u_j, u_j^*$ \ and \ $v_j, v_j^*$, in terms of the same basis $\{f_k |k\neq 0\}$, which makes the analysis cumbersome.
In order to avoid such a complication, we've intentionally chosen all functions to be real valued, in which case the matrix $R$ itself represents the Bogoliubov transformation:
\begin{equation}    \label{BdG_transform}
    V_{\hat{a}} = R \ V_{\hat{b}},
    \ \ 
    V_{\hat{a}} \equiv (.., \hat{a}_l^{\dagger}, \hat{a}_l, ..)^T,
    \ \
    V_{\hat{b}} \equiv (.., \hat{b}_l^{\dagger}, \hat{b}_l,..)^T.
\end{equation}
Here the vectors $V_{\hat{a}}$ and $V_{\hat{b}}$ consist of the creation and annihilation operators of all modes orthogonal to the condensate, the symbol $T$ denotes a transpose operation.
The structure of the $R$ matrix (\ref{R}) proves that the Bogoliubov transformation is an orthogonal mixing of different modes, while the creation and annihilation operators of each mode are mutually squeezed by the Lorentz rotation.
The mode mixing is caused by the nonzero off-diagonal overlapping integrals $\Delta_{jk}$, $j \neq k$. 

In principle, one may deal with the complex-valued wave functions. Then, the complex overlapping integrals $\Delta_{jk} = g N \int f_j^* \phi^2 f_k d^3{\bf r}$ form a Hermitian (instead of symmetric) matrix within the Bogoliubov-de Gennes matrix that corresponds to a unitary (instead of orthogonal) diagonalizing matrix \(||p_{jk}||\). 
However, the block structure of the matrices would involve blocks of a size larger than $2 \times 2$ and different from just the Lorentz rotations. 
In a result, a further computation of the characteristic function would become more cumbersome since it uses the block structure of the Bogoliubov transform matrix.

\section{BEC occupation statistics via\\ the characteristic function}

Let us calculate the characteristic function $\Theta (u) \equiv \text{Tr} \big( e^{iu\hat{N}_{\text{ex}}} \hat{\rho} \big)$ by means of the Wigner transform technique \cite{WignerReviewTatarskii,Balazs1984,WignerReviewScully,Schleich2001} which casts the operator-valued functions into the complex-valued functions and reduces calculating the trace to the integration over the whole phase space.
This method has been implemented in \cite{Englert2002} for a simpler case of the uniform BEC in a weakly interacting Bose gas.

Hereinafter, we assume that both the condensate wave function $\phi$ and the quasiparticle wave functions $(u_j,v_j)$ are chosen to be real valued.
This does not limit the generality since the Bogoliubov-de Gennes equations are linear and symmetric with respect to the complex conjugation, that is, for every solution $(u_j,v_j;E_j)$ there is a complex conjugate one $(u_j^*,v_j^*;E_j)$ with the same energy.

The forward and inverse Wigner transforms for an operator-valued function $F(\hat{a}^\dagger,\hat{a})$ of the creation and annihilation operators $\hat{a}^\dagger$ and $\hat{a}$ are given by the integrals:
\begin{equation}
\begin{split}
    &W(\alpha^*,\alpha) = \int e^{-\gamma \alpha^* + \gamma^* \alpha} \
        \text{Tr}   \left( 
                e^{\gamma \hat{a}^\dagger - \gamma^* \hat{a}}     
                    F(\hat{a}^\dagger,\hat{a}) 
                \right)
        \frac{d^2 \gamma}{\pi},
    \\
    \\
    &F(\hat{a}^\dagger,\hat{a}) =
        \int W(\alpha^*,\alpha) \ e^{-\gamma \alpha^* + \gamma^* \alpha} \
        e^{\gamma \hat{a}^\dagger - \gamma^* \hat{a}}
        \frac{ d^2 \gamma \ d^2 \alpha}{\pi^2}.
\end{split}
\end{equation}

The Wigner transform of the statistical operator $\hat{\rho}$, introduced in Eq.~(\ref{H}), is
\begin{equation}    \label{Wr}
\begin{split}
    W_{\rho} &= \prod_j \left( 2 \tanh\frac{E_j}{2T} \right)
        \exp {\left( -2 \beta^*_j \beta_j \tanh \tfrac{E_j}{2T}  \right)}
    \\    
    &= e^{-V_{\beta}^T B V_{\beta}} \prod_j \left( 2 \tanh\frac{E_j}{2T} \right);
    \ \
    V_{\beta}  \equiv (\ldots, \beta_s^*, \beta_s, \ldots)^T,
    \\
    B &= \left[ \begin{matrix}  
                \tanh\tfrac{E_1}{2T} \ \sigma_x & 0 & \ldots \\  
                0 & \tanh\tfrac{E_2}{2T} \ \sigma_x & \ldots \\
                \ldots &   \ldots & \ldots
            \end{matrix} \right],
    \ \ 
    \sigma_x =  \left[  \begin{matrix}  0 & 1 \\  
                                        1 & 0   \end{matrix}    \right].
\end{split}
\end{equation}
Here the complex variables $\beta_j^*$ and $\beta_j$ are associated with the quasiparticle operators $\hat{b}_j^\dagger$ and $\hat{b}_j$ and constitute the vector $V_{\beta}$; $B$ is a block-diagonal matrix of a symmetric quadratic form.

The Wigner transform of the operator $\exp \big( i u \hat{N}_{\text{ex}} \big)$ is
\begin{equation}    \label{WN}
\begin{split}
    W_{\Nex}   &= \prod_{k} \frac{2}{e^{iu}+1} 
                    \exp{ \left(2 \alpha^*_k \alpha_k \frac{e^{iu}-1}{e^{iu}+1} \right)} 
            \\        
            &=  \exp \left( \frac{e^{iu}-1}{e^{iu}+1} V_{\alpha}^T A V_{\alpha} \right) \prod_k \frac{2}{e^{iu}+1};
            \\
    V_{\alpha} &\equiv (\ldots, \alpha_s^*, \alpha_s, \ldots)^T,
    \quad
    A   =   \left[  \begin{matrix}  
                \sigma_x & 0 & \ldots   \\  
                0 & \sigma_x & \ldots   \\
                \ldots & \ldots & \ldots
            \end{matrix} \right].
\end{split}            
\end{equation}
Here the complex variables $\alpha_k^*$ and $\alpha_k$ are associated with the dressed particle operators $\hat{a}_k^\dagger$ and $\hat{a}_k$, introduced in Eq.~(\ref{Nex-aa}), and constitute the vector $V_{\alpha}$; $A$ is a block-diagonal matrix of another symmetric quadratic form.

Thus, the characteristic function $\Theta \equiv \text{Tr} \big( e^{iu \hat{N}_{\text{ex}}} \hat{\rho} \big)$ can be calculated as the following integral
\begin{equation}  \label{CF=WW}
    \Theta(u) = \int    W_{\Nex}\left(\{\alpha^*_j, \alpha_j\}\right) \times
                        W_{\rho}\left(\{\beta^*_j, \beta_j\}\right)
                    \prod_{j} \frac{d^2\beta_j}{\pi},
\end{equation}
where we chose $\beta_j^*$ and $\beta_j$ as the integration variables for definiteness and further convenience. (One may use $\alpha_j^*$ and $\alpha_j$, instead, or even any other set of variables representing the entire space of the integration.)

The only remaining step for calculating $\Theta(u)$ is linking the variables $\alpha^*_k$ and $\alpha_k$ with the variables $\beta^*_j$ and $\beta_j$.
It is easy to do in view of an important property of the Wigner transform highlighted in \cite{Englert2002}: The linear similarity transformations of the ladder operators in the function $F\big(\{\hat{a}_k^\dagger,\hat{a}_k\}\big)$ carry over to the variables in its Wigner transform $W\big(\{\alpha_k^*, \alpha_k\}\big)$. 
Hence, the Bogoliubov transformation of the operator-valued vectors (\ref{BdG_transform}) immediately casts to the following relation between the arguments of the Wigner transforms:
\begin{equation}
   V_{\alpha} = R V_{\beta},
\end{equation}
where the transformation matrix $R$ is given by Eq.~(\ref{R}).

Calculating the multidimensional Gaussian integral in Eq.~(\ref{CF=WW}) by means of a well-known formula involving an inverse square root of the determinant of the related symmetric quadratic form, we get the characteristic function in an explicit form via the block-diagonal matrices $A$, $B$ in Eqs.~(\ref{WN}), (\ref{Wr}) and the Bogoliubov transform matrix $R$ in Eq.~(\ref{R}): 
\begin{equation}    \label{pre_CF}
    \Theta(u) = \frac{ 
                    \prod_j \frac{2}{e^{iu}+1} \tanh \frac{E_j}{2T}  
                }{
                  \sqrt{ \ \det (i \ B - i \ \tfrac{e^{iu}-1}{e^{iu}+1} \ R^T A R) \ } \ 
                }  .  
\end{equation}
As a function of the variable $z = e^{iu}-1$ convenient for further calculations, it has the following canonical form:
\begin{equation}    \label{CF_Lambda}
\begin{split}
    &\Theta(z)  = \frac{ 
                    \prod_j \tanh \frac{E_j}{2T}  
                }{
                    \sqrt{\ \det(i \Upsilon ) \ } \
                } 
        =    
        \sqrt{  \prod_j\frac{\tanh  \frac{E_j}{2T}}{\Lambda_j^{[+]}} 
                    \times 
                \prod_{j'}\frac{\tanh \frac{E_{j'}}{2T}}{-\Lambda_{j'}^{[-]}} 
        },
    \\
    &\Upsilon(z) = B - \dfrac{z}{2} (R^T A R - B).
\end{split}   
\end{equation}
Here $\Lambda_j^{[\pm]}(z)$ are the eigenvalues of the matrix $\Upsilon(z)$ which tend to the values $\pm \tanh (E_j / 2T)$ at $z \to 0$, respectively.
The products run over all quasiparticle, excited states $j$, except the Goldstone excitation of the zero energy.

The result in Eq.~(\ref{CF_Lambda}) is directly suitable for numerical calculations of the condensate and noncondensate occupation probabilities, $\rho_0(N_0)$ and $\rho_{\text{ex}}(N_{\text{ex}})$, for an arbitrary profile of the trapping potential.
Indeed, for any given number $\nu$ of the excited modes chosen for a simulation of the characteristic function, the problem is reduced to just a standard computing of a ($2\nu \times 2\nu$)-matrix determinant.
The only possible problematic issue on this way is computing a large number of relevant eigenfunctions of the BEC-modified Schr$\ddot{\text{o}}$dinger equation (\ref{mod_Shr}) with a sufficient accuracy. The latter could be quite intricate in a general case of 3D traps without special symmetries.

The most convenient way to analyze the BEC occupation statistics (\ref{CF_Lambda}) is to study the generating cumulants
\begin{equation}
    \tilde{\kappa }_m= \left. \frac{\partial ^m \ }{\partial z^m}\ln \Theta \right|_{z=0}.   
\end{equation}
They are closely related to the central moments, including the mean value, the variance and the third moment defining the skewness $\gamma$:
\begin{equation}
    \la \hat{N}_{\text{ex}} \ra = \tilde{\kappa }_1,
    \quad
    \sigma^2 = \tilde{\kappa}_2 + \tilde{\kappa }_1,
    \quad
    \gamma \sigma^3 = \tilde{\kappa }_3 + 3 \tilde{\kappa }_2 + \tilde{\kappa }_1. 
\end{equation}
The formulas for the higher moments are given in \cite{PRA2010} via the Stirling numbers.

Remarkably, the result in Eq.~(\ref{CF_Lambda}) proves that each cumulant is essentially the sum of separate quasiparticle contributions, 
\begin{equation}    \label{kappa_decomp}
    \tilde{\kappa}_m= \frac{1}{2} \sum _{j} \left(\tilde{\kappa}_m^{[j+]}+\tilde{\kappa}_m^{[j-]}\right),
\end{equation}
associated with the eigenvalues $\Lambda_j^{[+]}(z)$ and $\Lambda_j^{[-]}$, respectively. 
Calculating the $m$-th cumulant requires computing the derivatives of the eigenvalues $\Lambda_j^{[\pm]}(z)$ with respect to the variable $z$ up to the $m$-th order.
Since these derivatives are determined by an infinitesimally small vicinity of the zero value of the variable $z$, we can calculate them by means of the Schr$\ddot{\text{o}}$dinger matrix perturbation theory. Namely, let us consider the matrix function $\Upsilon(z)$ at small values of $z$ as a sum of the unperturbed, base matrix $\Upsilon(z=0) = B$ and a  perturbation $\delta \Upsilon = R^TAR - B$ of a small magnitude $(-z/2)$.
The eigenvalues and associate eigenvectors of the unperturbed matrix are, respectively, $\pm \tanh (E_j/2T)$ and $e_{j\pm} = \tfrac{1}{\sqrt{2}}(\ldots,0,0,+1,\pm1,0,0,\ldots)^T$, where only the $j$-th pair of components is nonzero. In this basis, the perturbation matrices in Eq.~(\ref{CF_Lambda}) have the following entries
\begin{equation}  \label{AB}
\begin{split}
    &e_{j+}^T (R^TAR) \, e_{s-} = e_{j-}^T (R^TAR) \, e_{s+}  = 0,
    \\
    &A_{js}^{[+]} \equiv e_{j+}^T (R^TAR) \, e_{s+} 
        = +\sum_l p_{jl} p_{sl} e^{+\xi_{jl}+\xi_{sl}},
    \\
    &A_{js}^{[-]} \equiv e_{j-}^T (R^TAR) \, e_{s-} 
        = -\sum_l p_{jl} p_{sl} e^{-\xi_{jl}-\xi_{sl}},
    \\
    &e_{j+}^T B e_{s-} = e_{j-}^T B \, e_{s+}  = 0,
    \\
    &B_{js}^{[+]} \equiv e_{j+}^T B \, e_{s+} = + \delta_{js} \tanh \frac{E_j}{2T},
    \\
    &B_{js}^{[-]} \equiv e_{j-}^T B \, e_{s-} = - \delta_{js} \tanh \frac{E_j}{2T};
\end{split}
\end{equation}
$\delta_{jk}$ denotes a Kronecker delta.
The result for the perturbed eigenvalues of the matrix $\Upsilon(z)$ follows right away:
\begin{equation}    \label{D_Lambda}
\begin{split}
    &\Lambda_j^{[\pm]} \hspace{-1pt} = \hspace{-1pt}
        \pm \bigg[ \! \tanh \frac{E_j}{2T}
        \hspace{-1pt} - \hspace{-1pt}
                \frac{ a_j^{[\pm]} z }{2} 
        \hspace{-1pt} + \hspace{-1pt}
                \frac{ b_j^{[\pm]} z^2 }{4}
        \hspace{-1pt} - \hspace{-1pt}
                \frac{ c_j^{[\pm]} z^3 }{8} 
        \hspace{-1pt} + \hspace{-1pt}
                \text{O}(z^4)
        \bigg]\hspace{-1pt},
    \\
    &\quad  a_j^{[\pm]} = A_{jj}^{[\pm]} - \tanh\frac{E_j}{2T},
    \\       
    &\quad  b_j^{[\pm]} = \sum_{s\neq j} 
                \frac{\Big(A_{js}^{[\pm]}\Big)^2}
                {\tanh \frac{E_j}{2 T} - \tanh \frac{E_s}{2 T}},
    \\
    &\quad  c_j^{[\pm]}
                = \sum_{s\neq j}  \frac{\Big(A_{js}^{[\pm]}\Big)^2}
                {\tanh \frac{E_j}{2 T} - \tanh \frac{E_s}{2 T}}
                - \sum_{s\neq j}  \frac{A_{jj}^{[\pm]} \  \Big(A_{js}^{[\pm]}\Big)^2}
                {\left(\tanh \frac{E_j}{2 T} - \tanh \frac{E_s}{2 T}\right)^2}
    \\                    
     & \qquad \ \ + \sum_{s,r \neq j} 
            \frac{ A_{js}^{[\pm]} A_{sr}^{[\pm]}
                    A_{rj}^{[\pm]} }
            {   \left(\tanh \frac{E_j}{2 T} - \tanh \frac{E_s}{2 T}\right)
                \left(\tanh \frac{E_j}{2 T} - \tanh \frac{E_r}{2 T}\right)} .
\end{split}    
\end{equation}
Here, for simplicity's sake, we assumed that the eigenvalues are nondegenerate. Anyway, the degeneracy could be removed by some deformation of the trap.

Eq.~(\ref{D_Lambda}) yields all contributions to the first three cumulants in Eq.~(\ref{kappa_decomp}) explicitly:
\begin{equation}    \label{kappa_mode_mixing}
\begin{split}
    \tilde{\kappa}_1^{[j\pm ]} &=  
    \frac{ A_{jj}^{[\pm]} }{e^{E_j/T} - 1}
    + \frac{ A_{jj}^{[\pm]} - 1}{2},    
    \\[5pt]
    \tilde{\kappa}_2^{[j\pm ]} &= 
        \left( \tilde{\kappa}_{1}^{[j\pm]} \right)^2 
        - \left(\frac{ b_j^{[\pm]} }{ e^{E_j/T} - 1 } 
        + \frac{ b_j^{[\pm]} }{2} \right) \!,
    \\
    \tilde{\kappa}_3^{[j\pm ]} &= 
        2\left( \tilde{\kappa}_{1}^{[j\pm]} \right)^3 
    \!  - 3\tilde{\kappa}_{1}^{[j\pm]}
    \!      \left(\frac{ b_j^{[\pm]} }{ e^{E_j/T} - 1 } 
            + \frac{ b_j^{[\pm]} }{2} \right)
    \\
    &\ + \frac{3}{2} \left(\frac{ c_j^{[\pm]} }
                                { e^{E_j/T} - 1 } 
            + \frac{ c_j^{[\pm]} }{2} \right)\!.
\end{split}    
\end{equation}
The latter results are written in terms of the $A_{js}^{[\pm]}$ matrices in Eq.~(\ref{AB}), while the contributions from the $B_{js}^{[\pm]}$ matrices have been plugged in explicitly.
% [CHECK ALL CUMULANTS !!]
% \begin{widetext}
% \begin{equation}    \label{kappa_mode_mixing}
% \begin{split}
%     \tilde{\kappa} _1^{[j\pm ]} &= \frac{1}{2} \left(\frac{a_{j\pm}}{e^{E_j / T}-1}+\frac{a_{j\pm}-1}{2}\right),
%     \\
%     \tilde{\kappa}_2^{[j\pm ]} &= \frac{1}{2}\left(\frac{\Lambda_{j\pm}^{(1)}}{e^{E_j / T}-1}+\frac{\Lambda_{j\pm}^{(1)}-1}{2}\right)^2+ \frac{1}{2} \left(\frac{b^{[j\pm ]}}{e^{E_j / T}-1}+\frac{b^{[j\pm]}}{2}\right),
%     \\
%     \tilde{\kappa} _3^{[j\pm ]} &=
%         \left(\frac{a^{[j\pm ]}}{e^{E_j / T}-1}+\frac{a^{[j\pm ]}-1}{2}\right)^3+\frac{3}{2}\left(\frac{a^{[j\pm ]}}{e^{E_j / T}-1}+\frac{a^{[j\pm ]}-1}{2}\right)\left(\frac{b^{[j\pm ]}}{e^{E_j / T}-1}+\frac{b^{[j\pm ]}}{2}\right) 
%         \\
%         &-\frac{3}{4}\left(\frac{b^{[j\pm ]}}{e^{E_j / T}-1}+\frac{b^{[j\pm ]}}{2}\right)
%         +\frac{3}{4}\left(\frac{c^{[j\pm ]}}{e^{E_j / T}-1}+\frac{c^{[j\pm ]}}{2}\right)
% \end{split}
% \end{equation}
% \end{widetext}

It is worth noting that the representation of the characteristic function as a product of the eigenvalues involves a certain degree of freedom.
Namely, instead of the matrix $\Upsilon(z)$ and its eigenvalues, one could use another matrix $\tau_\text{left} \Upsilon(z) \tau_\text{right}$ and its eigenvalues provided the otherwise arbitrary matrices $\tau_\text{left}$ and $\tau_\text{right}$ possess a unity determinant. 
Our choice of the matrix $\Upsilon$ allows one to separate the contributions from the eigenvalues $\Lambda_j^{[+]}$ and $\Lambda_j^{[-]}$ as much as possible since, in terms of the Schr$\ddot{\text{o}}$dinger perturbation theory, the $[+]$-eigenvalue contributions don't introduce any corrections to the $[-]$-eigenvalue ones, and vice versa. 

The result (\ref{kappa_mode_mixing}) clearly reveals how the contributions from the thermal fluctuations (associated with the factor $1/(e^{E_j / T}-1)$ in the Boltzmann distribution of excitations over the energy levels) and contributions from the quantum fluctuations (associated with the nontrivial Bogoliubov transformation from the dressed particles to quasiparticles) combine together to constitute the overall cumulants. 
Namely, the mean occupation of the noncondensate (or condensate) is formed by additive contributions from the purely "thermal" and "quantum" depletions:
\begin{equation}
\begin{split}
    \la \Nex \ra = &\langle N_{\text{ex}}^{[T]}\rangle
    +
    \langle N_{\text{ex}}^\text{[qd]} \rangle, \\
    &\langle N_{\text{ex}}^{[T]}\rangle = \sum _{j,k}\frac{p_{jk}^2 \left(E_j^2 + \epsilon_k^2\right)}
    {2\epsilon_k E_j \left(e^{E_j / T}-1\right)},
    \\
    &\langle N_{\text{ex}}^\text{[qd]} \rangle = \sum_{j,k} \frac{ p_{jk}^2 \left(E_j - \epsilon_k \right)^2}{4 \epsilon_k E_j}.
\end{split}
\end{equation}
However, this is not true for the variance, skewness, excess and higher moments (cumulants) of the order $m \ge 2$, for which the mixed terms are always present in the sum along with the purely "thermal" and "quantum" terms.

In the general case, the formulas for the cumulants $\tilde{\kappa}_m$ quickly become very cumbersome with their increasing order $m$.
The complex common properties of the cumulants of the BEC occupation statistics for the interacting dilute gas in an arbitrary trap will be analyzed elsewhere.
Below we consider only the special trap set by Eq.~(\ref{U}) for which the analysis turns out to be relatively simple. 
\vspace{-1mm}

\section{The diagonal approximation for the quasiparticles and BEC statistics: BEC-modified Schr$\ddot{\text{O}}$dinger equation}

The simplification in the evaluation of the nonuniform condensate statistics (\ref{CF_Lambda}), which stems from the particular choice of the external potential (\ref{U}) as stated in the last paragraph of Sect.~I, is based on a remarkable "quasidiagonal" pattern of the quasiparticles existing in such a system.
Namely, each quasiparticle's wave function is formed mostly by a single eigenfunction of the BEC-modified Schr$\ddot{\text{o}}$dinger equation (\ref{mod_Shr}), so that approximately $u_j({\bf r}) \propto v_j({\bf r}) \propto f_j({\bf r})$.
Hence, the off-diagonal overlapping integrals $\Delta_{jk}$, Eq.~(\ref{Dij}), are so small that the matrix $||p_{jk}||$ diagonalizing the Bogoliubov-de Gennes problem~(\ref{BdG_mode_mixing}) is close to the identity matrix $I$ and the mode-mixing effects are negligible.
Under these conditions, the quasiparticle spectrum is approximately determined by the diagonal overlapping~integrals via the formula $E_j=\sqrt{\epsilon_j^2+2\Delta_{jj} \epsilon_j}$ standard for the  Bogoliubov theory.

Let us briefly outline how this "quasidiagonal" property manifests itself in the system under consideration.
First of all, the BEC-modified Schr$\ddot{\text{o}}$dinger equation (\ref{mod_Shr}) with the confining potential (\ref{U}) allows one to separate the Cartesian variables and factorize the overlapping integrals $\Delta_{jk}$ in Eq.~(\ref{Dij}) into the product of three one-dimensional integrals.
The corresponding partial equations along the transverse axes $y$ and $z$ possess the uniform partial ground states and sine or cosine partial excited modes due to the imposed periodic boundary conditions.
Thus, the solutions with different transverse structures can't contribute to the nonzero overlapping integrals and effect of coupling for the quasiparticles.

All the nonuniform effects in the system are encoded into the partial equation along the $x$ axis which is equipped with the Dirichlet (zero) boundary conditions.
Its ground state (condensate) wave function is given by the elliptic Jacobi function and varies from the half-period sine to an almost constant function (quickly decreasing to zero just in the narrow boundary regions) with the interaction constant $g$ increasing from zero to the larger values:
\begin{equation}    \label{sn}
\begin{split}
    &\phi(x) = \sqrt{\frac{p K(p)}{K(p)-E(p)}} \ 
                \text{sn} \left( \left.2 K (p) \frac{x}{L} \right| p\right),
    \\
    &L/\xi = \sqrt{8K(p)\big(K(p)-E(p)\big) \ }.
\end{split}
\end{equation}
Its parameters are determined by the complete elliptic integrals of the first and second kinds, $K(p)$ and $E(p)$, through the well-known healing length of the condensate 
\begin{equation}    \label{xi}
\xi = \sqrt{\hbar^2 L^3 \big/ (2 M g N) \ }.
\end{equation}
\begin{figure}[t]   
\includegraphics*[width=8cm]{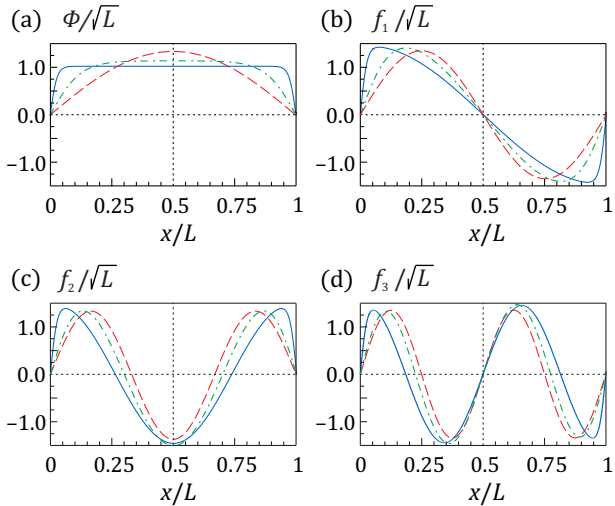}
\caption{   \label{fig-wave_func}
The profiles of (a) the condensate, $\phi(x)$, and (b)-(d) the first three dressed particle eigenfunctions, $f_j(x)$, of the $1$D version of the BEC-modified Schr$\ddot{\text{o}}$dinger Eq.~(\ref{mod_Shr}) along the axis $x$ with the Dirichlet (zero) boundary conditions for various interparticle interactions $g$ expressed via the ratio of the box trap size $L$ to the condensate healing length $\xi$, Eq.~(\ref{xi}): (i) an almost ideal gas, $L/\xi = 0.5$ (dashed red lines), (ii) a case of the moderate interaction, $L/\xi = 10$ (dot-dashed green lines), (iii) a case close to the Thomas-Fermi limit, $L/\xi = 50$ (solid blue lines). The two-component quasiparticle wave function $(u_j,v_j)$ has the same profile in view of the diagonal approximation of Sect.~V, $u_j(x) \propto v_j(x) \propto f_j(x)$.
} 
\end{figure}

\begin{figure}[t]
\includegraphics*[width=8.5cm]{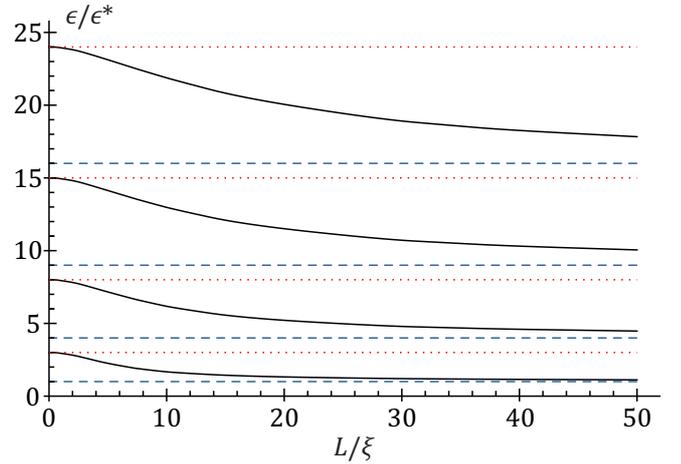}
\caption{The energy eigenvalues $\epsilon_j$ of the BEC-modified Schr$\ddot{\text{o}}$dinger equation (\ref{mod_Shr}) normalized to the energy scale $\epsilon^* = \hbar^2\pi^2/(2 M L^2)$ of the empty trap vs. the interparticle interaction expressed via the ratio of the box trap size $L$ to the condensate healing length $\xi$, Eq.~(\ref{xi}).
The solid lines, from the bottom to the top, correspond to the eigenfunctions which are uniform along the $y$ and $z$ axes, but have $1$, $2$, $3$ or $4$ zeroes along the $x$ axis.
The upper and lower bounds for each eigenvalue correspond to the ideal gas regime (dotted red lines) and the Thomas-Fermi limit (dashed blue lines), respectively.} 
\label{fig-energy}
\end{figure}
In the nearly ideal gas when the interparticle interaction is very small, $L/\xi \ll 1$, the profiles of the condensate and excited states are close to the sine functions $\sin \frac{\pi x}{L}$ and $\sin \frac{(j_x+1) \pi x}{L}$, where a nonnegative integer $j_x$ enumerates the longitudinal eigenfunctions. In this case, the longitudinal contribution to the dressed particle energy spectrum is $\epsilon_{j_x} \simeq \tfrac{\hbar^2\pi^2}{2 M L^2}\big((j_x+1)^2-1\big)$.
A significant restructuring of the condensate profile starts at $L/\xi \gtrsim 3$, when the condensate profile begins to flatten in the central part of the box trap and all, condensate and excited,  eigenfunctions acquire a steep varying region approaching the boundaries.
In the Thomas-Fermi limit, $L/\xi \gg 1$, the low-energy excitations are again close to the sine functions, however their phases are significantly shifted compared to the ideal gas case and the spectrum extends to the lower energies as $\epsilon_{j_x}\simeq \tfrac{\hbar^2\pi^2}{2ML^2}j_x^2$.
The high-energy eigenfunctions, which have many oscillations on the scale of the healing length $\xi$, have the usual quasiclassical quantization structure, but a further rise of the interaction, i.e., the $L/\xi$ ratio, transfers them into the group of the states with the properties similar to the aforementioned low-energy solutions.
A transition of a given state $j_x$ from the asymptotics of the $L/\xi\ll 1$ type to the asymptotics of the  $L/\xi \gg 1$ type is not uniform for different $j_x$, the low-energy excitations proceed much faster then the high-energy ones.
The entire picture of the eigenfunction/eigenvalue evolution with the increasing interaction outlined above is illustrated in Figs.~\ref{fig-wave_func},\ref{fig-energy}.

Numerical studies \cite{RPhys2019} have shown that the "quasidiagonal" property holds for the entire range of the interaction constant values and the contribution from the $j$-th bare particle to the $j$-th quasiparticle provides, at least, $97\%$ of the norm of the excited state wave function. 
This phenomenon of the persistently small values of the $1D$ overlapping integrals leading to the "quasidiagonal" pattern can be explained qualitatively as follows. 
In the case of an almost ideal gas, it is simply a consequence of the presence of a small parameter $g N$ in front of the integral in Eq.~(\ref{Dij}).
In the Thomas-Fermi limit, it is due to the facts that (i) the weight function under the integral (\ref{Dij}), $\phi^2$, is almost uniform and (ii) the solutions to the Schr$\ddot{\text{o}}$dinger equation (\ref{mod_Shr}) are truly orthogonal to each other with a constant weight. 
Finally, the intermediate range of the interaction constant values, squeezed between the other two ranges, appears to be quite narrow that prevents the overlapping integrals (which are constrained by the small-value asymptotics from the both sides) from reaching significant values.

The established "quasidiagonal" property hints at a factorization of the large matrix $\Upsilon(z)$, determining the charactaristic function (\ref{CF_Lambda}), into a direct product of $2\times 2$ blocks, i.e., at its ($2\times 2$)-block diagonal structure, while the evaluation related to each single block is easy to complete similar to \cite{PRA2000,Englert2002}.
However, the relation $||p_{jk}||\simeq I$ is only a necessary condition for the desired factorization. 
According to Eqs.~(\ref{D_Lambda}), (\ref{kappa_mode_mixing}), a factorized description of the cumulants is accurate enough only if one may keep just the first-order corrections $\sim a_j^{[\pm]}$ to the eigenvectors of the matrix representing the characteristic function and neglect all higher-order corrections ($\sim b_j^{[\pm]}$, $c_j^{[\pm]}$, etc.) responsible for the mode mixing.
So, the matrices $||p_{jk}^2 \frac{\epsilon_k}{E_j}||$ and $||p_{jk}^2 \frac{E_j}{\epsilon_k}||$ should also be close to the identity matrix~$I$.
The latter conditions are notably more restrictive than the original one, but should be required as well.

For the chosen system described in Sect.~II, all criteria listed above are fulfilled, which allows us to establish the following diagonal approximation for the BEC occupation statistics in terms of the characteristic function and generating cumulants of the probability distribution $\rho_{\text{ex}}$ for the total number of the noncondensed particles: 
\begin{equation}    \label{CF_diag}
\begin{gathered}
    \Theta(z)  = \prod_{j}
                \frac{1}{\sqrt{\left( 1-\zeta_j^{[+]} z \right)\left(1 - \zeta_j^{[-]} z \right)} 
        },
    \\    
    \tilde{\kappa}_m = \frac{(m-1)!}{2} \sum_{j} \left( \zeta_j^{[+]^m} + \zeta_j^{[-]^m} \right),       
    \\
    \zeta_j^{[\pm ]} = 
        \frac   {\left(E_j / \epsilon_j\right)^{\pm 1}}
                {e^{E_j/T}-1}+\frac{\left(E_j / \epsilon_j\right)^{\pm1}-1}{2}.
\end{gathered}    
\end{equation}
Here the sum and the product run over all quasiparticle states $j$ or, equivalently, over all excited states of the BEC-modified Schr$\ddot{\text{o}}$dinger equation (\ref{mod_Shr}).
The factors $\zeta_j^{[\pm]} \equiv a_j^{[\mp]} \big/ 2 \tanh (E_j/2T) = \tilde{\kappa}_1^{[j\mp]}$ introduced in \cite{RPhys2019} originate from the only non-negligible corrections $a_j^{[\pm]}$ retained in the expansions (\ref{D_Lambda}),~(\ref{kappa_mode_mixing}). This simplifies the expansion~(\ref{kappa_decomp}) as follows: $\tilde{\kappa}_m^{[j\pm]} = (m-1)! \big(\tilde{\kappa}_1^{[j\pm]}\big)^m$.

The result in Eq.~(\ref{CF_diag}) provides a complete description of the BEC-occupation statistics in the considered box trap (\ref{U}).
It has the same functional structure as the corresponding formulas found in \cite{PRA2000,Englert2002} for the uniform condensate.
The crucial difference is that now even the original particle states $(f_j({\bf r}); \epsilon_j)$ are not the bare particle modes and energies of the empty trap, but the eigenfunctions and eigenvalues of the BEC-modified Schr$\ddot{\text{o}}$dinger equation (\ref{mod_Shr}).
Simply put, in this theory not only the quasiparticles, but even the particles are dressed by the condensate.  
With the increase of the interaction constant $g$, they change significantly, as is shown in Fig.~\ref{fig-wave_func}. 
However, the "quasidiagonal" property still holds since the dressed particles and quasiparticles change almost proportionally to each other and, thus, the characteristic function always keeps its diagonal-approximation form of Eq.~(\ref{CF_diag}).
This important feature allows us to accurately account for the effects of the inhomogeneity and restructuring of the condensate and quasiparticle wave functions on the BEC fluctuations for the interactions ranging from zero in the ideal gas all the way to the values relevant to the Thomas-Fermi regime of BEC.

It is worth noting that, despite neglecting all dressed-mode mixing effects, the result in Eq.~(\ref{CF_diag}) is very different from the known result describing the BEC statistics in the ideal gas confined in an arbitrary trap \cite{PRA2014} or the result that one would get by formally plugging in the quasiparticle energies $E_j$ into the ideal-gas result instead of the bare particle energies $\epsilon_j$. 
In particular, Eq.~(\ref{CF_diag}) involves both the energy spectrum of the particles, $\{\epsilon_j\}$, and the energy spectrum of the quasiparticles, $\{E_j\}$. 
This is a consequence of the nontrivial Bogoliubov transformation between the particles and the quasiparticles and the fact that both the particles and the quasiparticles, introduced for the nonuniform condensate in Sect.~III, are dressed by the condensate in virtue of the BEC-modified Schr$\ddot{\text{o}}$dinger equation (\ref{mod_Shr}) and the Bogoliubov-de Gennes equations (\ref{GP+BdG}), respectively. 

In the simple case of the uniform condensate confined in the box trap with all periodic boundary conditions, the dressed particles and their spectrum $\{\epsilon_j\}$ are reduced to the usual bare particles and the spectrum of the empty box trap, respectively, so that the result in Eq.~(\ref{CF_diag}) is reduced to the known result obtained in \cite{PRA2000} for the usual Bogoliubov quasiparticles in the weakly interacting gas confined in the box. 
In the latter case, the overlapping integrals $\Delta_{ij}$ and Bogoliubov coupling between bare particles are restricted to just coupling within each pair of counter-propagating plane waves with the wave vectors ${\bf k}, {\bf -k}$, the matrix $\Upsilon$ becomes a $(2\times 2)$-block diagonal matrix, the $\text{det}(i\Upsilon)$ factorizes into a product of the determinants of the $2\times 2$ blocks, and the general formula in Eq.~(\ref{CF_Lambda}) reproduces its particular counterpart in \cite{PRA2000}.

Now, with the transparent picture and explicit formulas for the BEC statistics and cumulants (\ref{CF_diag}) as well as for the structure and spectrum of the dressed particles and quasiparticles presented above, we are ready to describe the BEC fluctuations in the nonuniform system at any level of the interparticle interactions.

\section{Main regimes of the BEC statistics}

We mainly focus on the most interesting, large picture of the BEC occupation statistics revealed in the large enough systems undergoing a transition from an ideal gas regime to a regime with a significant interparticle interaction. In this context, the system is large enough if a particle characteristic kinetic energy in the empty trap, 
\begin{equation}
    \epsilon^* = \hbar^2\pi^2\big/(2 M L^2), 
\end{equation}
is much less than the temperature, $\epsilon^* \ll T$. 
In this case, the BEC statistics is truly multi-mode statistics since there are plenty of excited states which may be well-occupied and thus significantly affect the fluctuations. 
In particular, this notion of a large system means that both the total number of the trapped particles and the thermal depletion of the condensate (at least, for weak interactions) are large in the absolute values: $N, \ \langle N_\text{ex}^{[T]} \rangle \gg 1$.

\begin{figure}[h]   
\includegraphics*[width=8.2cm]{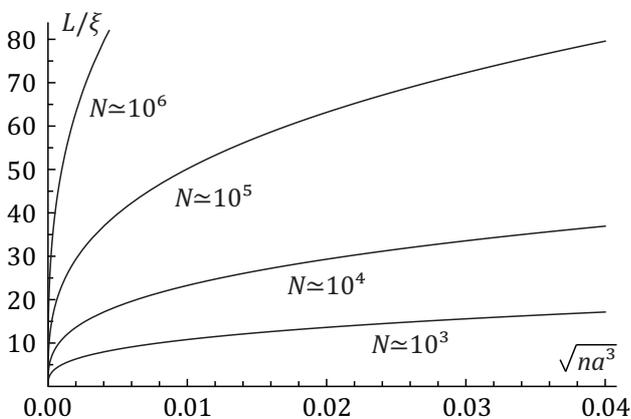}
\caption{   \label{fig-int-params}
The relation (\ref{na^3}) between two interaction parameters, $L/\xi$ and $\sqrt{n a^3}$, for different numbers of trapped particles $N$ characterising the size of the trap. 
The ratio $L/\xi$ measures the characteristic scale of the nonuniform condensate $\xi$ (the healing length (\ref{xi})) relative to the size of the trap $L$, and explicitly enters the Gross-Pitaevskii and Bogoliubov-de Gennes Eqs.~(\ref{GP+BdG}) according to the Eq.~(\ref{int-to-kin}). 
The parameter $\sqrt{n a^3}$ measures the $s$-wave scattering length $a$ relative to the average distance $n^{-1/3}$ between particles in the gas with density $n$. 
} 
\end{figure}

The effect of the interparticle scattering is determined by the ratio of the interaction and kinetic energy scales, 
\begin{equation}    \label{int-to-kin}
    g n / \epsilon^* = (L/ \pi \xi)^2 , 
    \end{equation}
where $\xi$ is the healing length (\ref{xi}). 
This ratio varies from zero in an ideal gas to quite large values in an interacting gas and may exceed the large parameter $T/\epsilon^*\gg 1$. 
According to Sect.~V, these parameters $L/\xi$ and $g n / \epsilon^*$ in Eq.~(\ref{int-to-kin}) are the most important and natural for describing the evolution of the BEC statistics with varying interparticle interaction since the structure and spectrum of the quasiparticles responsible for the BEC statistics exactly depend on them.
The other parameter, $n a^3$, which includes the $s$-wave scattering length $a$ and is widely used for characterizing the interaction, is related to $L/\xi$ as
\begin{equation}   \label{na^3}
\sqrt{n a^3} = \frac{(L/\xi)^3}{(8 \pi)^{3/2} N}.
\end{equation}
In fact, an extra large scaling factor $N \gg 1$ looks here rather artificial since it doesn't really matter for the structure of the condensate and quasiparticles. 
Just the product $g n$ enters the Gross-Pitaevskii and Bogoliubov-de Gennes equations (\ref{GP+BdG}) and, thus, is actually relevant. 
A relation between these two interaction parameters is illustrated in Fig.~\ref{fig-int-params} for experimentally relevant values.

We start the analysis of the BEC statistics regimes with evaluating the generating cumulants $\tilde{\kappa}_m$ in Eq.~(\ref{CF_diag}).
The $j$-th quasiparticle contribution to $\tilde{\kappa}_m$ is defined by the factors $\zeta_j^{[\pm]}$ and naturally split into the thermal, $\left(E_j / \epsilon_j\right)^{\pm 1} \big/ (e^{E_j/T}-1)$, and quantum, $\big((E_j / \epsilon_j)^{\pm1}-1\big)/2$, parts.
This leads to simply additive thermal and quantum terms for the mean condensate depletion $\tilde{\kappa}_1$.
For the higher-order cumulants $\tilde{\kappa}_m, m \ge2$, the picture is more complicated since the thermal and quantum contributions massively mixed due to the binomial expansion 
\begin{equation}    \label{zeta_binom}
    \zeta_j^{[\pm]^m} = \sum_{k=0}^m C_m^k \Bigg(\frac {\left(E_j / \epsilon_j\right)^{\pm 1}} {e^{E_j/T}-1}\Bigg)^k 
    \Bigg(\frac{\left(E_j/\epsilon_j\right)^{\pm1}-1}{2}\Bigg)^{m-k},  
\end{equation}
where $C_m^k = \frac{m!}{(m-k)!k!}$ is a binomial coefficient.

The leading terms of the thermodynamic-limit asymptotics of the pure thermal ($k=m$) or pure quantum ($k=0$) contributions to each cumulant in Eq.~(\ref{CF_diag}) from the entire spectrum of excitations are listed in Table~\ref{tab-contributions}.
It involves a numerical coefficient $R_{3/2}$ dependent on $gn/T$,
\begin{equation}    \label{R3/2}
    R_{3/2} = \frac{\pi}{2} \int_0^\infty
    \frac   {x^4 + \frac{gn}{T} x^2}
            {\sqrt{x^4 + 2 \frac{gn}{T} x^2}}
    \left(e^{\sqrt{x^4 + 2 \frac{gn}{T} x^2 }} - 1\right)^{-1} dx, 
\end{equation}
as well as numerical coefficients $S_m$ and $Q_m$ which give the scaled cumulant $\tilde{\kappa}_m$ in the corresponding thermally or quantum dominated asymptotics and depend on the interaction parameter $L/\xi$ through the dimensionless spectra of quasiparticles and dressed particles, $\{E_j/\epsilon^*\}$ and $\{\epsilon_j/\epsilon^*\}$.
For the ideal gas they reproduce the well-known results, in particular, $R_{3/2} = \pi^{3/2}\zeta(3/2)/8 \simeq 1.82$ where $\zeta(3/2)$ is a value of a Riemann zeta function $\zeta(q) = \sum_{k=1}^\infty k^{-q}$ at $q=3/2$; the other coefficients are discussed in the following subsections.

The order of magnitude of the cumulant $\tilde{\kappa}_m$ is determined by the largest of the two contributions presented in the Table~\ref{tab-contributions}, depending on the system parameters.
The scales of the mixed contributions are intermediate compared to the pure thermal and quantum ones and aren't shown explicitly.

\begin{table}   
\caption{   \label{tab-contributions}
The leading terms of the thermodynamic-limit asymptotics for the pure thermal and pure quantum contributions to the cumulants (\ref{CF_diag}) in terms of the temperature~$T$, particle's characteristic kinetic energy $\epsilon^* = \hbar^2\pi^2/(2 M L^2)$ and interaction energy $g n$.
}
\begin{ruledtabular}
\begin{tabular}{ccc}
cumulants
    & pure thermal part   
    & pure quantum part 
\\
\hline
\\
$\la \Nex \ra \equiv \tilde{\kappa}_1$
            & $R_{3/2} \left(\dfrac{T}{\epsilon^*}\right)^{\tfrac{3}{2}}$
            & $\dfrac{\sqrt{2} \ \pi}{12}
            \left( \dfrac{g n}{\epsilon^*}\right) ^{\tfrac{3}{2}}$
\\
\\
$\sigma^2 = \tilde{\kappa}_2+\tilde{\kappa}_1$
            & $ S_2 \left(\dfrac{T}{\epsilon^*}\right)^{2}$    
            & $ \dfrac{\sqrt{2} \ \pi^2}{16}
                \left( \dfrac{g n}{\epsilon^*}\right) ^ {\tfrac{3}{2}}$
\\
\\
$\tilde{\kappa}_3$
            & $ S_3 \left(\dfrac{T}{\epsilon^*}\right)^{3}$     
            & $ \dfrac{\sqrt{2} \ \pi}{16} \!
            \left( \dfrac{g n}{\epsilon^*}\right) ^{\!\tfrac{3}{2}} \!
           \ln  \dfrac{g n}{\epsilon^*}
            $
\\
\\
$\tilde{\kappa}_{m}, \ m \ge 4$
            & $S_m \left(\dfrac{T}{\epsilon^*}\right)^{m}$     
            & $Q_m \left( \dfrac{g n}{\epsilon^*}\right)^{\tfrac{m}{2}}$
            \\            
\end{tabular}            
\end{ruledtabular}
\end{table}

We calculated these leading terms of the asymptotics for a particular cumulant in the following way.
The sum in Eq.~(\ref{CF_diag}) has been converted into a continuous integral, if the integral is convergent.
In this case, the cumulant's order of magnitude is larger than the single energy-level contributions to it. 
In the opposite case, when the corresponding continuous integral is infrared-divergent because a relatively small group of low-energy excitations alone provides a major contribution to a particular cumulant, we employed the low-energy asymptotics $e^{E_j/T}-1 \simeq E_j/T$ and $E_j \simeq \sqrt{2\Delta_{jj}\epsilon_j}$ since the sum in Eq.~(\ref{CF_diag}) for such a cumulant quickly converges to its value already within the low-energy part of the spectrum. 
In the latter case, the individual quasiparticle contributions quickly decrease with the increase of the quasiparticle energy, so that the cumulant's value is of the same order of magnitude as the single low-energy-level contributions. 
A similar method had been previously applied to computing the BEC statistics in the ideal gas \cite{PRA2014,JPhysA2014}.

Table~\ref{tab-contributions} clearly shows that a leading role of the quantum or thermal contributions can't be directly decided upon a proximity of the system to the Thomas-Fermi limit. 
Indeed, approaching the Thomas-Fermi limit means making the healing length much smaller than the trap's size, $\xi \ll L$, which results into a large absolute value of the quantum contributions to the cumulants.
However, it says nothing about the relative values of the quantum and thermal contributions; in particular, the thermal contribution may prevail even in the Thomas-Fermi limit. 

Now we are ready to proceed with a description of the main regimes of the BEC statistics as they appear with increasing interparticle interactions.

\subsection{Thermally dominated, non-Gaussian fluctuations sensitive to the boundary conditions}

Let us consider first a so-called thermally dominated regime, which means that the interactions are quite weak and the quantum contributions are negligible compared to the thermal ones.
The corresponding BEC fluctuations are essentially non-Gaussian, as is expected for an ideal gas in a box trap with any boundary conditions \cite{PRA2010,PRA2014}, since, according to the Table~\ref{tab-contributions}, the variance is anomalously large compared to the mean value, $\sigma \propto \la \Nex \ra^{2/3} \gg \sqrt{\la \Nex \ra}$. 
In this case, the mean value is formed by a large number of excitations and it is the only cumulant which is much larger than the corresponding single energy-level thermal contribution.
At the same time, all higher-order cumulants, including the variance, are of the order of the single energy-level contributions. 
It means that the standardized probability distribution $\rho_\text{ex} (x)$ of the centered and normalized number of the noncondensed particles 
\begin{equation}  \label{x}
x=\big(N_{\text{ex}} - \la \Nex \ra \big) \big/ \sigma
\end{equation}
is characterized by the non-Gaussian higher-order cumulants $\tilde{\kappa}^{[x]}_m = \tilde{\kappa}_m/\sigma^m \propto (T/\epsilon^*)^0, \ m \ge 3$, which are nonzero even in the thermodynamic limit of a macroscopic system.
So, the BEC fluctuations are determined by a group of the low-energy quasiparticles, which is not large enough to make the central limit theorem work. Since the central limit theorem is not applicable here, the fluctuations are non-Gaussian and sensitive to modifications of a subgroup of these dominating excitations.
Such a modification occurs when the interaction constant $g$ varies so that the ratio $L/\xi$ is changed.

The values of the thermally dominated cumulants are
\begin{equation}  \label{Sm}
\begin{split}
    &\tilde{\kappa}_m = S_m \left(\frac{T}{\epsilon^*}\right)^m, \qquad m \ge 2;
    \\
    &S_m = \frac{(m-1)!}{2} \sum_j   
            \left[ \left( \frac{\epsilon^*}{\epsilon_j} \right)^m
            +
            \left( \frac{\epsilon^*}{\epsilon_j+2\Delta_{jj}} \right)^m \right].
\end{split}
\end{equation}
They follow from the asymptotics $\zeta_j^{[+]} \simeq T/\epsilon_j$ and $\zeta_j^{[-]} \simeq T/(\epsilon_j+2\Delta_{jj})$, which stem from the functional form of the quasiparticle spectrum $E_j=\sqrt{\epsilon_j^2 + 2\Delta_{jj}\epsilon_j}$  and provide the main thermal contributions in a large system.
Compared to the ideal-gas case \cite{PRA2000}, each cumulant looks like a mean of two ideal-gas cumulants calculated for the particle energy spectrum $\{\epsilon_j\}$ and the shifted spectrum $\{\epsilon_j + 2 \Delta_{jj}\}$.
When the interaction constant $g$ increases but the system is still in the thermally dominated regime of fluctuations, the cumulants and, hence, the probability distribution evolve in accord with the evolution of the eigenvalues $\epsilon_j$ and the rise of the overlapping integrals $\Delta_{jj}$ determined by the BEC-modified Schr$\ddot{\text{o}}$dinger equation (\ref{mod_Shr}).
Such an evolution, that may theoretically lasts with increasing interactions till the very Thomas-Fermi limit, $\xi \ll L$, and even further, qualitatively depends on the trap's boundary conditions and is determined by following two different trends.

First of all, the rise of the interaction constant $g$ significantly increases the overlapping integrals $\Delta_{jj}$, which tends to reduce the thermal contribution from $\zeta_j^{[-]}$ and doesn't affect the thermal contribution from $\zeta_j^{[+]}$.
For the noninteracting particles these contributions are equal to each other, while in the Thomas-Fermi limit, $L/\xi \to \infty$, the $\zeta_j^{[-]}$ contribution almost vanishes, which results into all anomalous cumulants becoming halved (compared to the case of the ideal gas).
The latter effect is, in fact, the squeezing of fluctuations, which has been studied in great detail in quantum optics \cite{Schumaker,Walls1994} and has been predicted for the uniform weakly interacting Bose gas in \cite{PRA2000}.

However, in the nonuniform system the structure and spectrum $\{\epsilon_j\}$ of quasiparticles nontrivially depend on the interparticle interactions, which makes the situation more interesting.
While the interaction is growing, the condensate becomes more and more flat along the $x$ axis and the excited energies $\{\epsilon_j\}$ notably decrease as is shown in Fig.~\ref{fig-energy}. For instance, the first excited energy level in the Thomas-Fermi limit is three times lower than that for the ideal gas (for details on the excitation spectrum, see \cite{RPhys2019}).
So, restructuring of the quasiparticles makes the excited states more accessible for the thermal population and tends to increase the $\zeta_j^{[+]}$ contribution to the thermally dominated anomalous cumulants.
This newly described effect has the same order of magnitude but the opposite sign as compared to the squeezing of fluctuations.

\begin{figure}[h]   
\includegraphics*[width=8.2cm]{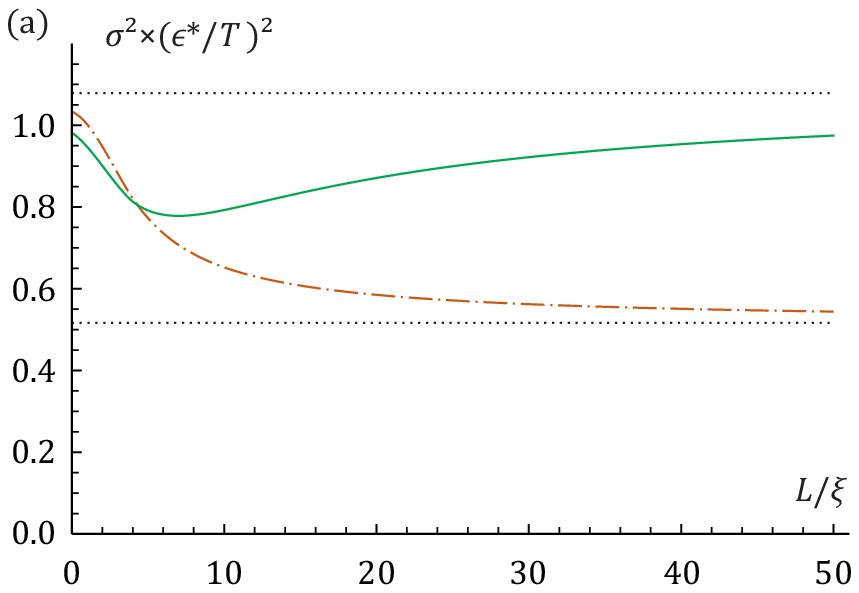}
\\[10pt]
\includegraphics*[width=8.2cm]{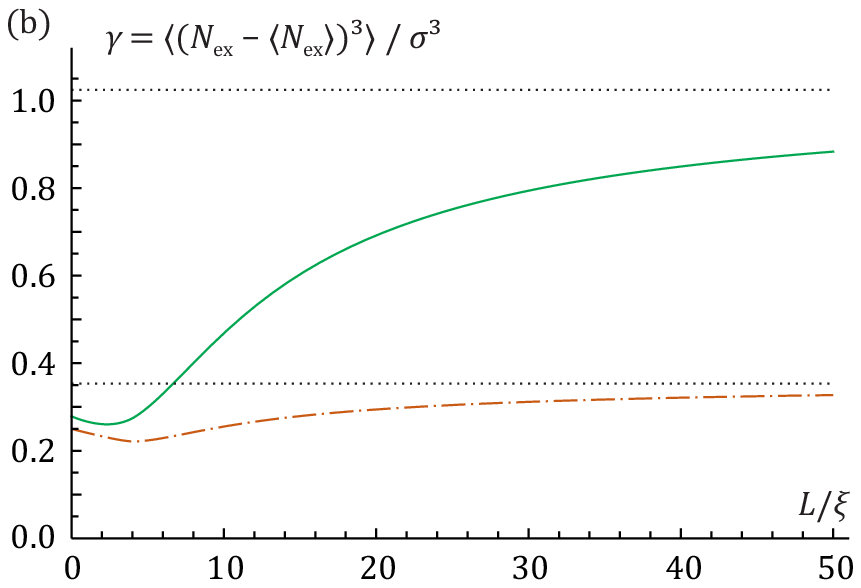}
\caption{   \label{fig-variance}    
(a) The scaled variance $\sigma^2$ and (b) the skewness 
$\gamma \equiv \langle \left( N_{\text{ex}} - \la \Nex \ra \right)^3\rangle / \sigma^3$ of the noncondensate-occupation probability distribution 
vs. the interaction parameter $L/\xi$ in a regime of the thermally dominated fluctuations. The "squeezing of fluctuations" effect in the box with all-periodic boundary conditions (the dot-dashed orange lines) is complemented by the effect of the restructuring of quasiparticle wave functions in the box with the Dirichlet boundary conditions along one of the axes (the solid green lines).
The dotted lines correspond to the Thomas-Fermi limit, $L/\xi \gg 1$, assuming that the system is still in a regime with prevailing thermal contributions.
} 
\end{figure}

The competition of these two effects is shown in Fig.~\ref{fig-variance} for the main thermal contributions to the variance $\sigma^2$ and skewness $\gamma$, characterizing the width and asymmetry of the noncondensate-occupation probability distribution, respectively.
We find these main, thermodynamic-limit thermal contributions from Eq.~(\ref{Sm}) as follows
\begin{equation}
    \sigma^2 \to S_2 \left( \frac{T}{\epsilon^*}\right)^2,
    \
    \gamma \equiv \frac{\la(\Nex - \la \Nex \ra)^3\ra}{\sigma^3} \to \frac{S_3}{S_2^{3/2}}.    
\end{equation}

For the relatively weak interactions, $L/\xi \lesssim 1$, the density profile of the condensate is almost the same as that in the ideal gas and, hence, the squeezing of fluctuations is the dominant effect. 
It means that the variance is decreasing when the interparticle scattering is getting more intense.
The condensate wave functions starts to change rapidly when the interaction grows to such an extent that $L/\xi \gtrsim 3$. Starting from this value of the interaction, the effect of the quasiparticle restructuring comes into play.
Since the contributions from the $\zeta_j^{[-]}$ are already mostly suppressed, a further increase in the interaction constant $g$ makes the restructuring effect dominant. 
As a result, the main thermal contribution to the variance grows and, close to the Thomas-Fermi limit, becomes even larger than that for the ideal gas.
For the higher-order cumulants the situation is qualitatively similar as is illustrated by the graph of the skewness (the distribution's asymmetry determined by the third cumulant) in Fig.~\ref{fig-variance}.

\begin{figure*} 
\includegraphics[width=17cm]{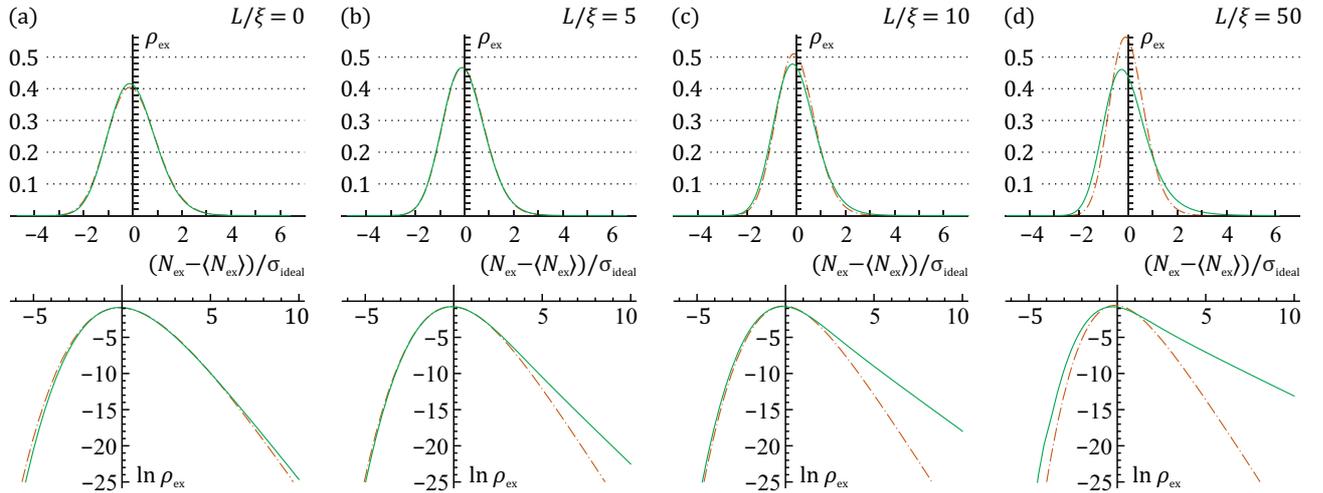}
\caption{ \label{rho(x)}
Thermally dominated probability distribution $\rho_{\text{ex}}$ (upper row) and its logarithm (lower row) as the functions of the scaled noncondensate occupation $(\Nex - \la \Nex \ra) / \sigma_\text{ideal}$ for a large enough Bose system confined in a box with the Dirichlet boundary conditions along the $x$ axis and periodic boundary conditions along the $y$ and $z$ axes (the solid green lines) or with all-periodic boundary conditions (the dot-dashed orange lines) for different interparticle interactions: (a) $L/\xi = 0$ (an ideal gas), (b) $L/\xi = 5$, (c) $L/\xi = 10$, (d) $L/\xi = 50$. 
The scaling employs the variance $\sigma_\text{ideal}$ for the ideal gas in the box with all-periodic boundary conditions, while the centering is done individually for each distribution.
The evolution of $\rho_\text{ex}$ is quite different in the cases of the uniform and nonuniform condensates.
The effect of the boundary conditions becomes more pronounced in the Thomas-Fermi limit $L/\xi \gg 1$.
} 
\end{figure*}

The evolution of the cumulants described above is in strong contrast to that in the case of the uniform interacting gas which enables only the effect of the squeezing of fluctuations alone.
Changing the boundary conditions, which introduces an inhomogeneity into the system, significantly affects the statistics of the BEC fluctuations as is shown in Fig.~\ref{rho(x)}.
This effect is especially pronounced for strong enough interactions when $L/\xi \gg 1$.
Naively speaking, one should not expect it since the healing length is small and the boundary conditions are somewhat screened.
However, the consistent analysis of the quasiparticles and statistics reveals, via Eqs.~(\ref{CF_diag}), the crucial for this effect fact -- due to the nontrivial Bogoliubov transform from the dressed particles to quasiparticles and back, the information about the spectrum of the dressed particles does not disappear from the cumulants.

\subsection{The crossover from the anomalous \\ to Gaussian statistics}

The first quantum statistical effect which becomes significant with the increasing interparticle interaction is the quantum depletion of the condensate.
It comes into play when the quantum contribution to the mean noncondensate occupation becomes as large as the thermal depletion, $\la N_{\text{ex}}^{[T]}\ra \simeq \la N_{\text{ex}}^\text{[qd]}\ra$, that is, when the interaction energy becomes on the order of the temperature, 
\begin{equation}  \label{crossover-mean}  
\frac{g n}{\epsilon^*} \sim \frac{T}{\epsilon^*}, \quad \textrm{or} \quad
\sqrt{ n a^3} \sim \frac{3\sqrt{\pi}}{8} \left(\frac{T}{T_c}\right)^{3/2}.
\end{equation}
(Note that at this point the thermal depletion is about $0.4$ of that in the ideal gas due to a decreased value of the coefficient $R_{3/2}$ presented in Table~\ref{tab-contributions}.)
At the same time, at this interaction a significant presence of the quantum depletion is the only difference from the regime of the thermally dominated fluctuations since, according to Table~\ref{tab-contributions}, all higher-order cumulants, including the variance, remain thermally dominated.
It means that the BEC-occupation statistics remains non-Gaussian, dependent on the low-energy excitations and the variance is still anomalously large.

More changes occur when, due to a further increase in the interaction, the quantum contribution to the variance turn out to be of the same importance as the thermal one.
A corresponding threshold interaction is on the order of
\begin{equation}  \label{crossover-variance}
    \frac{g n}{\epsilon^*} \sim \left(\frac{T}{\epsilon^*}\right)^{4/3}\!\!,
    \ \ \textrm{or} \ \ 
    \sqrt{n a^3} \sim 
    \frac{8 S_2 N^{1/3}}{\pi^{5/2} \zeta^{4/3}\big(\frac{3}{2}\big)} \! \left(\frac{T}{T_c}\right)^2 .
\end{equation}
Assuming $S_2 \simeq 1$, the latter threshold could be written as $\sqrt{n a^3}\sim 0.13 N^{1/3}(T/T_c)^2$. (The actual dependence $S_2$ on $L/\xi$ is shown in Fig.~\ref{fig-variance}a.)
At this stage, the mean noncondensate occupation is specified mostly by the quantum depletion and the BEC fluctuations are no longer anomalous. 
The standard deviation starts to scale as is typical of the usual Gaussian fluctuations in the macroscopic thermodynamics, $\sigma \propto \sqrt{\la \Nex \ra}$.
At the same time, all cumulants ${\tilde \kappa}_m$ of the higher orders $m\ge3$ are still thermally dominated, and their scaling is still consistent with that of the variance (despite its value has been increased by the quantum contributions).
Thus, the fluctuations are still described by a non-Gaussian standardized distribution $\rho_\text{ex} (x)$ of the stochastic variable in Eq.~(\ref{x}) which remains asymmetric, although the variance is already of a standard magnitude.
This is a marginal case separating Gaussian and non-Gaussian regimes of the BEC statistics. 
%$\zeta(\frac{3}{2})$ is a value of a Riemann zeta function $\zeta(x)$ at $x=\frac{3}{2}$.

The two-step crossover from the anomalous BEC statistics to the Gaussian one outlined above assumes that the interaction parameter specified in Eq.~(\ref{crossover-mean}) should be smaller than that in Eq.~(\ref{crossover-variance}).
The latter requirement implies a large enough thermal contribution to the condensate depletion, since the second threshold of the interaction parameter $\sqrt{na^3}$, Eq.~(\ref{crossover-variance}), is larger than the first one, Eq.~(\ref{crossover-mean}), by an extra factor $N^{1/3} (T/T_c)^{1/2}$ determined by the thermal depletion of the ideal gas $\sim N(T/T_c)^{3/2}$.
This formal condition is satisfied starting from small system sizes corresponding to $10-100$ thermally excited atoms on average in the ideal gas regime.
Such minimal noncondensed ensembles are so small that their BEC statistics is not the thermodynamic-limit one, but a non-Gaussian statistics in either quantum dominated or thermally dominated regimes because of the mesoscopic, finite-size effects.

Another restriction for a validity of the picture described above is related to the fact that the entire analysis is done within the Bogoliubov approximation, which is valid for $\sqrt{na^3} \lesssim 0.04$.
The crossover happens if the temperature of the system is quite low, so that both thresholds in Eqs.~(\ref{crossover-mean}), (\ref{crossover-variance}) lie within the specified range of the interactions.
Say, for $T/T_c \simeq 0.15$ the threshold interaction (\ref{crossover-mean}) is $\sqrt{na^3} \simeq 0.04$, meaning that in this case the entire crossover to the quantum dominated statistics occurs at the interactions exceeding the border of validity of the Bogoliubov approach.

\subsection{The quantum dominated, Gaussian fluctuations}

A further rise of the interaction finally turns the BEC statistics into the standard Gaussian law of fluctuations typical of the macroscopic thermodynamics. 
This happens when the quantum contributions prevail in the variance -- at the interactions much larger than in Eq.~(\ref{crossover-variance}), 
\begin{equation}    \label{quantum-regime}
    \frac{g n}{\epsilon^*} \gg \left(\frac{T}{\epsilon^*}\right)^{4/3}\!\!,
    \ \ \textrm{or} \ \
    \sqrt{n a^3} \gg 
    \frac{8 S_2 N^{1/3}}{\pi^{5/2} \zeta^{4/3}\big(\frac{3}{2}\big)} \left(\frac{T}{T_c}\right)^2\!\!.
\end{equation}
The quantum dominated variance is effectively accumulated over a very wide spectrum of excitations, which is in contrast to the case of the thermally dominated regime in which the narrow infrared portion of the quasiparticle spectrum dominates the variance. 
As a result, the central limit theorem becomes applicable to the system 
and the variance scales as a normal, Gaussian standard deviation, $\sigma \propto \sqrt{\la \Nex \ra}$.
This circumstance significantly increases the magnitude of the variance in comparison to the higher-order cumulants.
From the scaling laws of the Table~\ref{tab-contributions}, it is easy to infer that the cumulants of the higher orders $m\ge3$ are surpassed by the quantum dominated variance --- no matter which contributions, quantum or thermal, make these cumulants up. Besides, the strong effect of the boundary conditions is lost.

The Gaussian-type fluctuations are usual and, probably, the most expected for the statistical physics of the macroscopically large many-body systems.
Thus, it is even more nontrivial that, for the Bose condensate in the three-dimensional box traps, the existence of the Gaussian, quantum dominated regime of the BEC statistics is, in fact, essentially a mesoscopic effect.
Strictly speaking, this regime can't survive in the system and, hence, disappears in the thermodynamic limit.
Indeed, let us consider what happens if the system size $L$ is increasing while the density of the gas $n$, temperature $T$ and the interaction constant $g$ (as well as the scattering length $a$) are kept constant.
Then the parameters characterizing the thermal and quantum contributions according to Table~\ref{tab-contributions} both increase proportionally to $L^2$,
\begin{equation}
    \frac{T}{\epsilon^*} \propto \frac{g n}{\epsilon^*} \propto L^2. 
\end{equation}
The thermal contribution to the variance $\propto(T/\epsilon^*)^2$ grows faster than the quantum one $\propto\left(g n/\epsilon^*\right)^{3/2}$.
Even if we start with a quantum dominated regime, enlarging the system sooner or later definitely makes the thermal contributions to the variance prevailing, and the thermally dominated regime unavoidably arises: The thermodynamic-limit statistical distribution of the total number of condensed/noncondensed particles becomes non-Gaussian and characterized by the anomalously large (compared to the mean value) fluctuations.
The only quantum effect that does not vanish in the thermodynamic limit, $L\to \infty$, is the quantum depletion of the condensate since the fraction of the "depleted" particles also goes as the volume of the trap $\propto L^3$, like the thermal depletion and the BEC-condensate occupation.

The same conclusions are obvious also in terms of the parameters $\sqrt{na^3}$ and $T/T_c$ since then the $N$ is the only parameter which varies under proceeding to the standard thermodynamic limit.
Thus, the corresponding inequality in Eq.~(\ref{quantum-regime}) will definitely be violated with an increase of the total number of particles $N$. 

\subsection{An interplay between different regimes \\in a mesoscopic system}

\begin{figure}
\includegraphics[width=8.3cm]{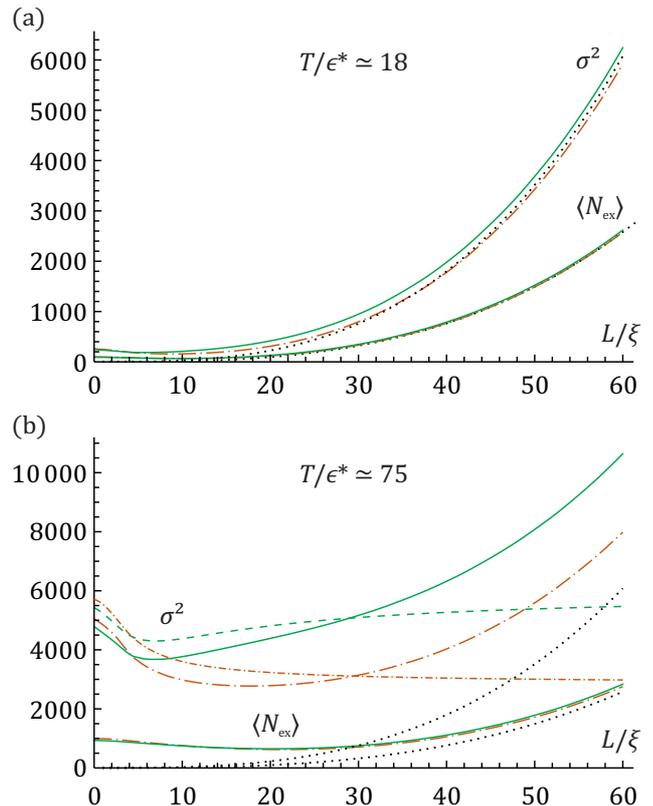}
\caption{   \label{fig-sigma_and_Nex}    
The variance $\sigma^2$ and mean value $\la \Nex \ra$ of the noncondensate occupation vs. the interaction parameter $L/\xi$, Eq.~(\ref{xi}), in the mesoscopic box trap with the boundary conditions which are periodic either along all axes (the dot-long-dashed orange lines) or just along two axes, but with the Dirichlet (zero) boundary conditions along the third axis (the solid green lines).
The size of the mesoscopic trap corresponds to the parameter $T/\epsilon^*$ equal (a) $18$ or (b) $75$, that approximately amounts to $100$ or $1000$ noncondensed particles $\la \Nex \ra$ in the ideal gas limit $L/\xi \to 0$.
The dotted lines represent the asymptotics of the quantum dominated regime of fluctuations.
The thermally dominated regime, taking place at $L/\xi \lesssim 15$ for (a) and at $L/\xi \lesssim 30$ for (b), is almost washed out for the smaller system (a), while it is quite notable for the larger system (b). The evolution of the variance in the thermally dominated regime roughly goes after the dashed green or dot-short-dashed orange lines representing the asymptotics of the scaled cumulant $S_2 (L/\xi)$ for the two different boundary conditions and calculated in Fig.~\ref{fig-variance}(a) in the thermodynamic-limit approximation.
} 
\end{figure}

\begin{figure}
\includegraphics[width=8.5cm]{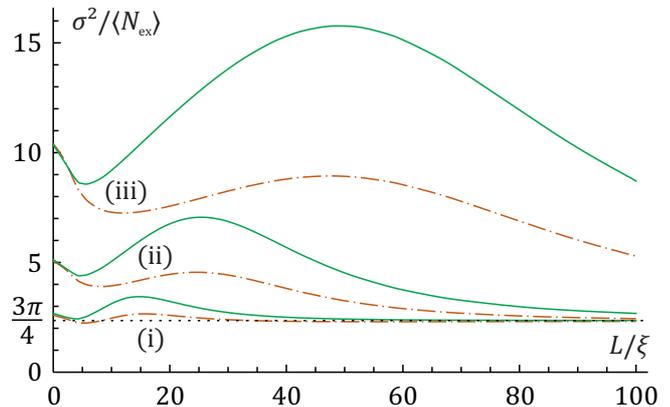}
\caption{   \label{fig-sigma_to_Nex}  
The ratio of the variance and the mean occupation of the noncondensate, $\sigma^2/\la \Nex \ra$, vs. the interaction parameter $L/\xi$, Eq.~(\ref{xi}), in the mesoscopic box trap with the boundary conditions which are periodic either along all axes (the dot-dashed orange lines) or just along two axes, but with the Dirichlet (zero) boundary conditions along the third axis (the solid green lines).
Different pairs of the lines are for different sizes of the trap corresponding to the parameter $T/\epsilon^*$ equal (i)~$18$, (ii)~$75$ or (iii)~$325$ that approximately amounts to $100$, $1000$ or $10000$ noncondensed particles $\la \Nex \ra$ in the ideal gas limit $L/\xi \to 0$.
The dotted line corresponds to the asymptotics $\sigma^2/\la \Nex \ra = 3\pi/4$ describing the standard Gaussian statistics in a regime of the quantum dominated fluctuations. 
With increasing size of the mesoscopic system, the region of the thermally dominated fluctuations is extending to larger values of the interaction parameter $L/\xi$ and the pronounced intermediate region of the maximal, most "anomalous" relative variance of the noncondensate fluctuations appears. 
}
\end{figure}

Yet, the quantum contributions to the BEC fluctuations could be well pronounced since the experiments are dealing with the mesoscopic systems.
As a result, there is an interesting interplay between the different regimes of the BEC statistics outlined above as is illustrated in Figs.~\ref{fig-sigma_and_Nex} and \ref{fig-sigma_to_Nex}. 
We mostly address the variance of the condensate (or noncondesate) occupation fluctuations since the higher-order cumulants demonstrate a similar behavior, and also because the scaling of the variance, in fact, determines the regime of the BEC fluctuations.

Fig.~\ref{fig-sigma_and_Nex} clearly demonstrates how the quantum contributions arising with increasing interaction parameter $L/\xi$ truncate the thermally dominated regime at some value of $L/\xi$ (about 15 and 30 for the plots (a) and (b), respectively)  which depends on the size of the system. 
We consider the case $T/\epsilon^* \gg 1$ when the absolute value of the variance can be approximated by a sum
\begin{equation}  \label{interplay}
    \sigma^2 \simeq S_2(L/\xi) \left( \frac{T}{\epsilon^*} \right)^2 
             + \frac{\sqrt{2}}{16\pi} \left(\frac{L}{\xi}\right)^3 ,
\end{equation}
in which just the main, pure thermal and pure quantum contributions are kept. The omitted terms have a smaller magnitude; in particular, the mixed quantum and thermal term contributes a summand of the order of $(T/\epsilon^*)^{3/2}$.
While we turn on and increase the interaction, the evolution of the variance in the mesoscopic system first qualitatively follows the evolution of the coefficients $S_2$ in Eq.~(\ref{Sm}) and the BEC statistics is non-Gaussian. Then it gives a way to a different process, which is a transition to the quantum dominated, Gaussian regime of the BEC statistics, and the coefficient $S_2$ no longer determines the value of the second cumulant.
If the system is large enough, as in Fig.~\ref{fig-sigma_and_Nex}(b), the truncation happens at a relatively large value of $L/\xi$, so the thermally dominated regime extends over a significant part of the evolution path of the $S_2$ coefficient. 
In this case, the effect of the boundary conditions possibly rises to the values which are significant even in the regime of the quantum dominated BEC statistics which becomes dominant at the larger values of the interaction parameter $L/\xi$.
In the opposite case of a small system, as in Fig.~\ref{fig-sigma_and_Nex}(a), the thermally dominated regime occurs only in a relatively narrow range of the small values of the interaction parameter $L/\xi$. In this case, the entire thermally dominated regime in the evolution of the variance could be washed-out, since the restructuring of the condensate and quasiparticles starts when the dominating contributions to the BEC statistics are already quantum.

The interplay between the statistical regimes could be further clarified by considering the relative variance of the BEC fluctuations, namely, the ratio of the variance and the occupation of the noncondensate, $\sigma^2/\la \Nex \ra$, which is shown in Fig.~\ref{fig-sigma_to_Nex} as a function of the interaction parameter $L/\xi$.
It is convenient to rewrite the approximation (\ref{interplay}) for the standard deviation in the form
\begin{equation}  \label{interplay-2}
    \sigma^2 \simeq 
        \frac{S_2}{R_{3/2}^{4/3}}
                \langle N_\text{ex}^{[T]} \rangle^{4/3}
        + \frac{3\pi}{4} \langle N_\text{ex}^\text{[qd]} \rangle
\end{equation}
explicitly referring to the thermal, $\langle N_\text{ex}^{[T]} \rangle$, and quantum, $\langle N_\text{ex}^\text{[qd]} \rangle$, parts of the condensate depletion as per Table~\ref{tab-contributions}.
It makes obvious that the increasing interaction moves the system towards the regime of the Gaussian fluctuations characterized by the ratio $\sigma^2/\la \Nex \ra = 3\pi/4$. 
If the thermal fraction of the depletion is rather small, as is the case when the system is small or cold enough, the ratio $\sigma^2/\la \Nex \ra$ is nowhere significantly different from this limiting value, and the thermally dominated regime is not pronounced.
However, if the thermal fraction of the depletion is significant, the situation is different: The thermally dominated variance at small values of the interaction parameter $L/\xi$ is anomalously large in comparison with the mean value, $\sigma^2/\la \Nex \ra = \frac{S_2}{R_{3/2}}\sqrt{T/\epsilon^*}$.
In the latter case, the additional, potentially huge factor $\sqrt{T/\epsilon^*} \gg 1$ makes the anomalous scaling of the variance quite distinguishable from the opposite, quantum dominated regime.
This difference may be observed in the experiments and employed as an indicator which determines whether the system is in the thermally dominated regime of the BEC statistics or quantum dominated one.

The results of the calculations for the mesoscopic system shown in Fig.~\ref{fig-sigma_to_Nex} also reveal another interesting feature in the dependence of the ratio $\sigma^2/\la \Nex \ra$ on the interparticle interaction. 
Namely, between the initial decrease of this ratio caused by the squeezing of fluctuations at small values of the interaction parameter $L/\xi$ and the final decrease to the Gaussian-limit value $3\pi/4$ due to the quantum effects, there is an intermediate region of a pronounced increase of this ratio.
This peak is located slightly to the left of the lower boundary, Eq.~(\ref{crossover-mean}), of the crossover region of the BEC statistics, where a standard deviation starts to grow due to the quantum contributions while the mean noncondensate occupation is not yet rising and even decreasing. 
The latter happens despite the quantum depletion has already been coming into play, as is predicted by the two-step crossover picture described in Sect.~VI.B, because the rise of the quantum depletion is slower than the decrease of the thermal depletion governed by a decrease of the coefficient $R_{3/2}$ with increasing ratio $gn/T$.
The restructuring of the condensate and quasiparticles is very important for this effect since, as is shown in Fig.~\ref{fig-sigma_to_Nex}, the effect is much stronger for the nonuniform condensate, that is, for the trap with the Dirichlet (zero) boundary conditions along one of the axes. Remarkably, the maximal, most "anomalous" relative value of the BEC fluctuations, $\sigma^2/\la \Nex \ra$, is achieved in the crossover regime of the BEC statistics when the system is mesoscopic, that is, still far from the thermodynamic, macroscopic limit.

%Clearly, for $T \sim \ldots$ the quantum and thermal contributions to the mean value  $\langle N_{ex} \rangle$ appears to be of the same order, which somewhat suppress the anomalous relation between the mean value and the variance. (CHECK!!!)
%The crossover between the non-Gaussian, thermally dominated regime and the Gaussian, quantum dominated regime happens when contributions to the variance of different natures are of the same order. 
%In such conditions, the mean value $\la \Nex \ra$ is mostly determined by the quantum depletion, which means the BEC fluctuaions are characterised by standard-order variance, $\sigma \propto \sqrt{\la \Nex \ra}$.
%However, at the same time all higher cumulants are mostly determined by the thermal contributions and thus the standardized probability distributions remains asymmetric and non-Gaussian.
%Further increase of the interaction constant leads the system towards the quantum-dominated regime, where the thermal contributions are negligible. 
%According to (\ref{!}), this regime means a standard dispersion,  $\sigma \propto \sqrt{\la \Nex \ra}$ and the Gaussian standardized distribution $\rho_x$ in the thermodynamic limit, since the scaled cumulants vanish while the ratio $g \langle n_0 \rangle / \epsilon^*$ goes to infinity.

%%%%%
%%%%%
%%%%%

\section{Conclusions and prospects for the experimental studies}

To summarize, we consistently describe the statistics of the total number of condensed (or noncondensed) particles in a dilute Bose-condensed gas of the weakly interacting particles trapped in the box with two different boundary conditions, either periodic along all axes or Dirichlet (zero) along one of the axes. 
We develop the analytical theory disclosing this statistics for the {\it nonuniform} condensate within the mean-field approach and apply it for computing the quasiparticle spectrum, profile and particle content as well as parameters of the BEC-occupation probability distribution as the functions of the interparticle interaction. 
In this way, we find how the BEC-occupation statistics depends on the interparticle interaction and boundary conditions, both for the mesoscopic and macroscopic systems. These results are consistently derived in the present paper from the first principles and agree with the preliminary analysis \cite{Entropy2018} that had been based on an ad hoc model assumption about the spectrum and structure of the quasiparticles. 

We reveal two qualitatively different regimes of the BEC statistics, the thermally dominated non-Gaussian and quantum dominated Gaussian ones, as well as describe the crossover between them.
Remarkably, we disclose a delicate, nontrivial transition from an ideal-gas regime of the negligible interactions to the Thomas-Fermi regime of the pronounced interactions and nonuniformity of the condensate. 
This transition is driven by two competing effects, the squeezing of the BEC fluctuations due to the Bogoliubov coupling and the restructuring of the quasiparticles due to increasing nonuniformity of the condensate. The latter effect hadn't been discussed in the relevant literature, mostly limited to the analysis of the BEC fluctuations for the case of the uniform condensate. 

For a strong enough interparticle scattering, the quantum effects and, in particular, the quantum depletion prevail over the thermal ones.
This leads to the so-called quantum dominated regime of the BEC statistics, which means that the probability distribution of the number of condensed particles tends to the normal, Gaussian distribution standard for the macroscopic thermodynamics (the variance is of the order of the mean value, $\sigma^2 \propto N$).

In the opposite case of relatively weak interactions, the thermal contributions to the BEC fluctuations are prevailing and the statistical regime becomes thermally dominated.
In the latter regime, the BEC-occupation fluctuations acquire the anomalous scaling (for the box traps, one has $\sigma^2 \propto N^{4/3}$), and the BEC-occupation probability distribution remains essentially non-Gaussian and dependent on the boundary conditions even in the thermodynamic limit.
We find the crossover region between these two regimes and show that in the actual mesoscopic traps the thermally dominated regime could be achieved in a wide range of parameters overlapping with the Thomas-Fermi BEC regime. 

We explain the boundary-conditions effect on the BEC fluctuations in the thermally dominated regime by a violation of the central limit theorem: The main contributions to the BEC fluctuations come from relatively few low-energy excitations which are quite sensitive to the boundary conditions imposed on the trap.
The effect of the boundary conditions exists even in the case of the relatively strong interparticle interactions sufficient to stipulate the Thomas-Fermi approximation, that is, in the case when the healing length of the condensate is much smaller than the size of the trap, $\xi \ll L$, and so the boundary effects would seem to be shielded.

Let us remind that the results presented in Sect.~V,~VI are heavily based on the special, quasidiagonal pattern of the quasiparticle states taking place in the flat traps with the Dirichlet (zero) boundary conditions along one of the axes.
Namely, each quasiparticle wave function is mainly formed by just one eigensolution to the BEC-modified Shcr$\ddot{\text{o}}$dinger equation (\ref{mod_Shr}). This fact makes the analysis of the BEC fluctuations much simpler since (a) the general matrix representation of the characteristic function in Eq.~(\ref{pre_CF}) is reduced to the compact analytical formula in Eq.~(\ref{CF_diag}) and (b) finding each quasiparticle requires, as per Sect.~V, computing just one eigensolution to the Shcr$\ddot{\text{o}}$dinger equation (\ref{mod_Shr}).
For comparison, the analogous problem of finding the first longitudinal excitation in a can-like trap, implemented in a straightforward way in \cite{Hadzibabic2019excitations}, involved solving a multiple mode iterative problem each step of which has been accompanied by orthogonalizing a basis of the employed functions with respect to the condensate profile. Such a procedure could hardly be implemented analytically -- only numerically.

For an arbitrary trapping potential profile the quasidiagonal pattern of the quasiparticle states is not guaranteed. Moreover, in many cases it is not valid.
Hence, the evolution of the BEC statistics with increasing interparticle interactions, in the general case, involves another manifestation of the quasiparticle restructuring. This is changing the quasiparticle composition associated with the coupling of the dressed particles into the quasiparticles due to an appearance of the considerable off-diagonal overlapping integrals $\Delta_{jk}$ and deviation of the matrices $||p_{jk}||$ and $R$, diagonalizing the Bogoliubov-de Gennes problem as per Eqs.~(\ref{BdG_mode_mixing})--(\ref{R}), from their diagonal approximation adopted in Sect.~V.
This quasiparticle-composition part could, for some traps different from the one considered above, contribute to the effect of the restructuring of quasiparticles on the BEC statistics at the same order of magnitude as the part that is analyzed in the present paper and stems from the restructuring of the dressed particles' spectrum and spatial profiles.  
The aforementioned contribution is beyond the diagonal approximation and will be discussed elsewhere.

Nevertheless, the model analyzed in the present paper allows us to overview the known experiments relevant to the BEC statistics in the flat traps and discuss some prospects for further experimental studies of various regimes and features of the BEC statistics.
 
In a majority of such experiments, the thermal fluctuations play a very significant part and the truly quantum dominated regime has never been achieved yet.
Consider a very interesting relevant experiment on the BEC in an almost uniform trap \cite{Hadzibabic2017QD} dedicated to a quantitative measurement of the quantum depletion of the condensate. 
The system confines about $55000$ atoms in a cylindrical trap at the temperature about $T_c/6$, that corresponds to the thermal contribution of about $150$ atoms to the standard deviation $\sigma$. (At these conditions in the ideal gas regime, the mean number of the excited atoms would be about $3000$.)
Even for the largest interaction achieved in this experiment, $\sqrt{n a^3} \sim 0.035$, the quantum contribution to the standard deviation is about $100$ atoms, which brings the system only to the crossover region, where the thermally dominated regime just starts its transformation into the quantum dominated regime.

At the same time, the trap in the experiment \cite{Hadzibabic2017QD} is quite large and populated, so a wide range of parameters corresponding to the thermally dominated regime of the BEC statistics is already accessible by means of just rising the temperature.
It means that the nontrivial thermally dominated regime of the BEC fluctuations is achievable at the current level of technology and its features outlined in Sect.~VI above, including the effect of the quasiparticle restructuring due to the restructuring of the nonuniform condensate, could be tested in further experiments.
In fact, the first longitudinal excitation of the BEC-condensed gas confined in such a can-like trap has already been observed \cite{Hadzibabic2019excitations} -- its evolution with increasing interaction qualitatively coincides with the picture given in Sect.~V (compare Fig.~\ref{fig-wave_func} above with Fig.~3 from  \cite{Hadzibabic2019excitations}). 

Dependence of the BEC fluctuations on the condensate non-uniformity could be studied also by directly controlling the trapping potential $U_{\text{tr}}({\bf r})$. In particular, for a cylindrical optical trap one could employ the doughnut Laguerre-Gaussian laser beams \cite{Kugo1997} with different azimuthal indices $l$ that would allow to control a power-law steepness of the radial profile of the trapping potential, $U_{\text{tr}}(r) \propto r^{2l}$. The steepness of the longitudinal potential profile could be controlled by varying the intensity profile of the end-cap laser beams. 

Another relevant example of a flat trap is an optical box trap demonstrated in \cite{RaizenBECinBOx-PRA2005}. The number of $^{87}\text{Rb}$ atoms it confines, $N \lesssim 3500$, is controlled by evaporation timing and spacing, $d < 80 \ \mu m$, of the optical end-caps. It provides the BEC-condensed gas clouds corresponding to a wide range of controllable interparticle interactions enabling studies both the weak mean-field and strong Tonks-Girardeau regimes of BEC. An integrated atom-counting setup based on a fluorescence imaging technique with single atom detection capability allows one to determine the number of trapped atoms exactly. Such a setup looks very promising for studying the BEC statistics.  

The observation of both, thermally dominated and quantum dominated, statistical regimes and the crossover between them could be based on the simultaneous measurements of the mean numbers of the condensed and noncondensed particles, $\la N_0 \ra$ and $\la \Nex \ra$, as well as the variance $\sigma^2$.
The latter is rather challenging. 
However, the experimental techniques for measuring the second moment of the BEC-occupation statistics are already appearing (in particular, see \cite{Kristensen2019}) and, hopefully, will become accurate and reliable quite soon. 
The ratio $\sigma^2 / \la \Nex \ra$ as a function of the interaction parameter $L/\xi$ has been already discussed in Sect.~VI, Eq.~(\ref{na^3}) and Fig.~\ref{fig-sigma_to_Nex}.
Below, we present analysis of the ratio
$\sigma^2 / \la N_0 \ra$ and $\sigma^2$ which could help to distinguish between the two, non-Gaussian and Gaussian, statistics. 
We'll employ the interaction parameter $\sqrt{na^3}$ which could be convenient to measure and control in the experiments.

\begin{figure}[H]   
\includegraphics[width=8.5cm]{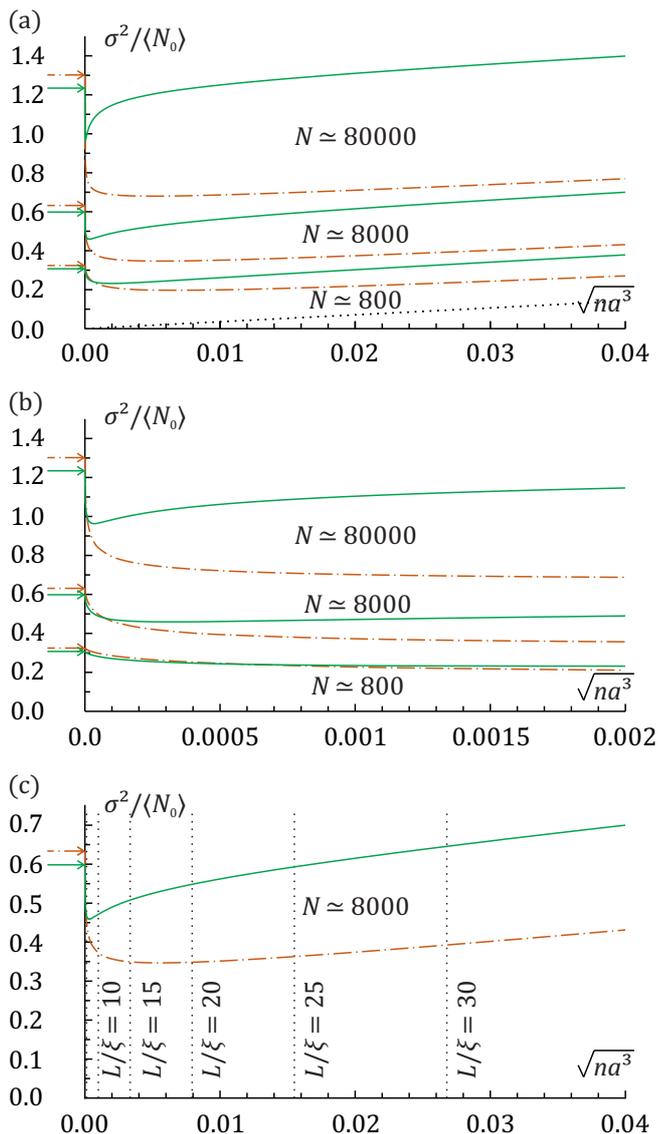}
\caption{   \label{fig-sigma^2-to-N0}
(a) The ratio of the variance and the mean occupation of the condensate, $\sigma^2/\langle N_0\rangle$, vs. the interaction parameter $\sqrt{n a^3}$ at the temperature $T = T_c/4$ for the different total numbers of particles $N = 800$, $8000$ or $80000$ confined in the mesoscopic box trap with the boundary conditions which are periodic either along all axes (the dot-dashed orange lines) or just along two axes, but with the Dirichlet (zero) boundary conditions along the third axis (the solid green lines).
The dotted line corresponds to the pure quantum contribution to the variance represented by the leading term of its thermodynamic-limit asymptotics as per Table~\ref{tab-contributions}.
The solid and dot-dashed arrows indicate the values of $\sigma^2/\langle N_0\rangle$ in the ideal gas.
(b) The enlarged part of (a) at small interactions, $\sqrt{na^3} < 0.002$, showing that the system is quickly going away from the ideal gas regime with increasing interparticle interaction.
(c) The enlarged part of (a) showing the graphs just for the case of $N = 8000$ and the values of the interaction parameter $L/\xi$ corresponding to the current values of $\sqrt{na^3}$.
} 
\end{figure}

Let us turn to Fig.~\ref{fig-sigma^2-to-N0} presenting the variance scaled by the mean condensate occupation, $\sigma^2 / \la N_0 \ra$, as a function of the interaction parameter.
Technically, this figure is based on the same set of the numerical data as we used in Fig.~\ref{fig-sigma_to_Nex}. 
Here we interpret this data as describing the mesoscopic system with approximately $800$, $8000$ or $80000$ particles trapped at the temperature $T \simeq T_c/4$. (The latter implies that in the ideal gas regime the mean number of the excited particles is $100$, $1000$ or $10000$, respectively.)
In all three cases the evolution from thermally dominated regime of fluctuations (smaller $\sqrt{na^3}$) towards the quantum dominated  regime (larger $\sqrt{na^3}$) is clearly seen and well pronounced.

At very small interactions, the scaled variance $\sigma^2 / \la N_0 \ra$ sharply drops.
Then, a significant growth of the scaled variance of the BEC occupation in the box with the Dirichlet (zero) boundary conditions along one of the axes (the solid lines) starts when the interaction parameter reaches a certain small value which depends on the number of trapped particles, namely, $\sqrt{na^3} \sim 0.00003$, $0.0003$ or $0.002$ in the case of $N \simeq 80000, 8000$ or $800$, respectively. 
It is a manifestation of the effect of the restructuring of the quasiparticles due to restructuring of the nonuniform condensate, discussed in Sect.~VI, which results in a considerable deviation of the variance from the decreasing variance of the uniform condensate fluctuations in the box with all periodic boundary conditions shown by the dot-dashed lines in Fig.~\ref{fig-sigma^2-to-N0}. 
The point is that this effect is absent for the uniform condensate which is subject to just the effect of the squeezing of fluctuations due to the Bogoliubov coupling. 
The latter always tends to decrease the variance of fluctuations. 
Finally, at the larger values of the interaction parameter $\sqrt{na^3}$, the variance demonstrates notable linear growth, which signifies the transition to the quantum dominated regime.

This picture stems from Eqs.~(\ref{interplay}), (\ref{interplay-2}) written as 
\begin{equation}    \label{sigma^2-to-N0}
    \frac{\sigma^2}{\la N_0 \ra} \simeq 
\frac{16 S_2 N^{1/3}}{\pi^2 \zeta^{4/3}(3/2)} \left( \frac{T}{T_c} \right)^2 + 
2 \sqrt{\pi na^3} .
\end{equation}

The linear trend at large $\sqrt{na^3}$ values is driven by the second, quantum summand tending to the quantum dominated regime of the BEC statistics with the normal Gaussian fluctuations. 
For a given value of the interaction parameter $\sqrt{na^3}$, the quantum dominated regime is described by the linear dependence of the variance of fluctuations on the number of the condensed particles, $\sigma^2 \simeq 2 \sqrt{\pi n a^3} \langle N_0 \rangle$, which is typical for the fluctuations in the macroscopic thermodynamics. 
The dominance of the first, thermal summand, which includes a large extra factor $N^{1/3} \gg 1$, corresponds to the thermally dominated regime of the non-Gaussian, anomalous fluctuations.

The thermally dominated regime leads to a much more complicated picture -- the left part of Fig.~\ref{fig-sigma^2-to-N0}a, the details of which are enlarged in Figs.~\ref{fig-sigma^2-to-N0}b,c.
The peak of the variance scaled by the mean condensate occupation, $\sigma^2 / \langle N_0 \rangle$, within the thermally dominated region is achieved in the ideal gas limit, that is, at the zero interaction, $\sqrt{na^3} =0$, and is indicated by the solid or dot-dashed arrow in Fig.~\ref{fig-sigma^2-to-N0}. 
Such a scaled variance doesn't have a second peak in the crossover region, contrary to the variance scaled by the mean noncondensate occupation, $\sigma^2 / \la \Nex \ra$, which demonstrates a second, higher and wider peak shown in Fig.~\ref{fig-sigma_to_Nex}. 
The reason is that the mean number of the condensed particles in the first scaling factor, $N_0 \approx N$, is almost constant and close to the total number of trapped particles for all temperatures $T \ll T_c$ considered above, while the mean number of the excited particles in the second scaling factor, $\la \Nex \ra$, significantly depends on the interparticle interactions.

Note, that for the chosen parameters of the system, the quantum regime itself is not achieved: The maximum value of the interaction parameter $\sqrt{na^3} \simeq 0.04$, included in Fig.~\ref{fig-sigma^2-to-N0}, corresponds to the ratio $L/\xi \simeq 16, 34$ or $74$ for the total number of the trapped atoms $N \simeq 800, 8000$ or $80000$, respectively, which refers just to the first step of the crossover. 
Thus, the thermal contribution to the variance is prevailing in the entire range of the curves shown in Fig.~\ref{fig-sigma^2-to-N0} so that the constant shift between the mesoscopically calculated curves and the dotted line, representing the quantum regime asymptotics, remains, along with the effect of the boundary conditions, pronounced for all values of $\sqrt{na^3}$ in Fig.~\ref{fig-sigma^2-to-N0}(a).
Nevertheless, all the nontrivial picture of the crossover is already clearly seen from Fig.~\ref{fig-sigma^2-to-N0} since the subsequent monotonic approach to the quantum dominated regime with increasing interactions looks straightforward. 
Anyway, there is no fundamental restriction for achieving the pure quantum regime of the BEC fluctuations. It requires only the lower temperatures and more populated trap.
In particular, for the ensembles of $9000$, $90000$ or $900000$ trapped bosons at the temperature $T = T_c/20$ (so that, again, there are approximately $100$, $1000$ or $10000$ excited particles on average in the ideal gas regime) the value of the interaction parameter $\sqrt{na^3} \simeq 0.04$ corresponds to the ratio $L/\xi \simeq 35, 75$ or $170$, respectively, which is larger than the $L/\xi$ values for both boundaries of the crossover regime specified in Eqs.~(\ref{crossover-mean}), (\ref{crossover-variance}). 

\begin{figure}[H]   
\includegraphics[width=8.2cm]{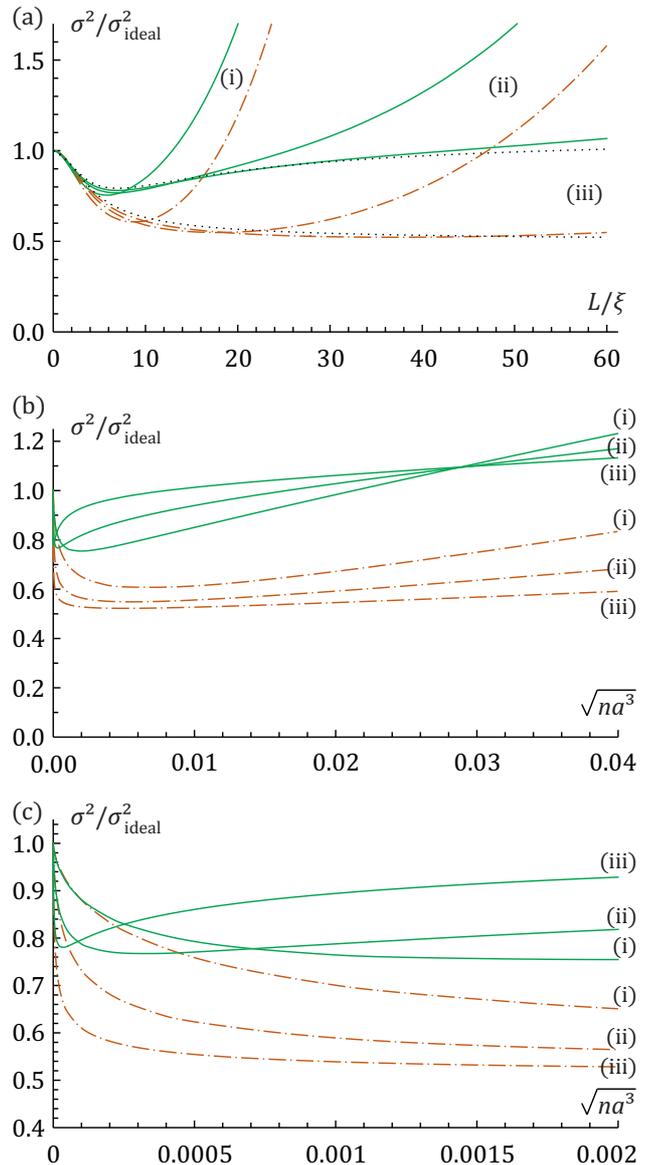}
\caption{   \label{fig-sigma^2-to-ideal}
The variance of the BEC-occupation fluctuations $\sigma^2$ scaled by its value in the ideal gas $\sigma^{2}_{\text{ideal}}$ in the mesoscopic box trap with the boundary conditions which are periodic either along all axes (the dot-dashed orange lines) or just along two axes, but with the Dirichlet (zero) boundary conditions along the third axis (the solid green lines), vs. the interaction parameter: (a) $L/\xi$, Eq.~(\ref{xi}), or (b) and (c) $\sqrt{na^3}$.
The three pairs of the lines (i), (ii), and (iii) correspond to the three different total numbers of trapped particles $N = 800$~(i), $8000$~(ii) or $80000$~(iii) that amounts to the noncondensate occupation $\la \Nex \ra \simeq 100$, $1000$ or $10000$ in the ideal gas limit $L/\xi \to 0$ at $T=T_c/4$.
The dotted lines correspond to the asymptotics of the variance in the thermally dominated regime, $\sigma^2 \simeq S_2(L/\xi) (T/\epsilon^*)^2$, shown in Fig.~\ref{fig-variance}, Table~\ref{tab-contributions} and discussed in Sect.~VI.A.
The larger the mesoscopic system (i.e., the number of the trapped particles $N$), the wider is the range of the interaction parameter $L/\xi$  where the BEC statistics follows the thermally dominated regime of the non-Gaussian fluctuations. The part (c) magnifies the details of a quick departure of the system from the ideal gas regime with increasing interparticle interactions.
} 
\end{figure}

The experimental measurement and interpretation of the absolute value of the condensate/noncondensate variance as a function of the interaction parameter $\sqrt{na^3}$ or $L/\xi$ requires a graph with a different, absolute scaling. Let us choose it to be the variance, $\sigma^2_{\text{ideal}}$, of the BEC-occupation fluctuations for the ideal gas trapped in the box with the corresponding boundary conditions, either periodic along all axes or Dirichlet along one of the axes. This ideal-gas variance is fully understood and calculated for all temperatures, including the critical region \cite{PRA2010,PRA2014,JStatPhys2015}.  
Such a scaled variance as the function of the interaction parameter $L/\xi$, presented in Fig.~\ref{fig-sigma^2-to-ideal}(a), shows that the larger the system the larger is the interaction until which the variance follows the asymptotics $\sigma^2 \simeq S_2(L/\xi) (T/\epsilon^*)^2$ describing the thermally dominated regime of the BEC statistics and determined by the scaled cumulant $S_2$, Eq.~(\ref{Sm}). The same variance as the function of the other interaction parameter, $\sqrt{na^3}$, related to $L/\xi$ by Eq.~(\ref{na^3}) (Fig.~\ref{fig-int-params}) is presented in Fig.~\ref{fig-sigma^2-to-ideal}(b). The details of its fast going away from the ideal gas regime at small interactions are enlarged in Fig.~\ref{fig-sigma^2-to-ideal}(c).

Figs.~\ref{fig-sigma^2-to-N0} and \ref{fig-sigma^2-to-ideal} clearly show that designing the experiments on measuring the BEC fluctuations and demonstrating different regimes of the BEC fluctuations implies finding a compromise between two opposite requirements.
First of all, to distinguish between the two, quantum dominated and thermally dominated, statistical regimes on the basis of the scaled variance $\sigma^2/\langle \Nex \rangle$ or $\sigma^2/\langle N_0 \rangle$ one should deal with sufficiently large system, since the thermal contribution to the variance is larger than the quantum one only by a cubic root of the total number of particles, $\sim N^{1/3}$. The latter should be large enough to clearly observe the difference between the two regimes.
At the same time, if one pretends to observe the nontrivial competition between the squeezing of fluctuations due to the Bogoliubov coupling and the effect of the quasiparticle restructuring, the low-value range of the interparticle interaction parameter $0 < L/\xi < 20$ should be achieved and well resolved in the experiment.

Fulfilling both requirements at the same time seems challenging since, for the Bose gases with the large total number of trapped atoms $N \gg 1$, the dependence of the parameter $L/\xi$ on the interaction parameter $\sqrt{na^3}$, directly controllable in the experiment, is rather fast as is shown in Fig.~\ref{fig-int-params}. So, the experiment should deal with very small and still well controllable values of the interaction parameter $\sqrt{na^3}$ (say, near the zero crossing in a Feshbach resonance) that could make difficult to access the whole interval of small interactions, $0 < L/\xi < 5$, where the squeezing of fluctuations dominates the variance's behavior in the systems with the nonuniform condensate. 
Yet, the measurements at the interactions $L/\xi \sim 10$ look quite accessible for reasonably large systems with the to-

\noindent tal number of particles $N \sim 10^4$.
In the latter case, the competition between the various effects, predicted above and illustrated in Fig.~\ref{fig-variance}, could be verified experimentally.
For instance, measuring a positive derivative of the thermally dominated variance with respect to the interaction parameter would definitively mean observing the effect of quasiparticle restructuring on the BEC statistics due to restructuring of the nonuniform condensate.

A direct measurement of the effect of the boundary conditions on the BEC fluctuations is also accessible. 
For example, one could employ a trap which confines the Bose gas inside a toroidal volume (such as a toroidal all-optical trap experimentally demonstrated in \cite{toroidBEC-PRL2011}) with an external potential in the shape of a wall produced by a laser radiation and crossing the toroid in the radial direction. By controlling such a potential wall, one could control the azimuthal boundary conditions in the toroidal trap, that is the same as switching the boundary conditions at the top and bottom of the can-like flat trap \cite{Hadzibabic2019excitations}. In particular, they could be switched from the periodic (no wall) to Dirichlet (a very high potential wall) boundary conditions. 
It is possible also to mimic switching between different boundary conditions directly in the can-like trap by means of adding an extra potential wall in the central part of the can, which would be permeable for atoms but affect the particle wave functions existing in the trap.
As is explained in Sect.~IV, affecting the lower-energy well-populated particle states, for instance, by turning the extra wall on or off, leads to the same effect on the BEC fluctuations as changing the boundary conditions.

Finally, let us briefly comment on the protocols and procedures for measuring the BEC-occupation statistics and sampling the fluctuating numbers of particles in the condensate and noncondensate, $N_0$ and $\Nex$. In principle, the most direct and reliable method to measure the variance, or the second moment, of the condensate or noncondensate occupation fluctuations is to employ measurements of some physical quantity associated with the corresponding quadratic operator $\hat{N}_{0}^2$ or $\hat{N}_{\text{ex}}^2$. Then, averaging over the ensemble of the experimental shots would unambiguously give the corresponding second moment of the BEC statistics, $\langle \hat{N}_0^2 \rangle =\text{Tr}(\hat{N}_0^2 \hat{\rho})$  or $\langle \hat{N}_{\text{ex}}^2 \rangle = \text{Tr} (\hat{N}_{\text{ex}}^2 \hat{\rho})$, defined by the statistical operator (\ref{H}) for a particular value of the interparticle interaction parameter in Eq.~(\ref{na^3}) which is controllable in the experiment, say, by tuning the Feshbach resonances \cite{Chin2010,Kohler2006}. 

Besides, the sum of the two stochastic variables $N_0$ and $N_{\text{ex}}$, which is equal to the total number of particles confined in the trap, $N_0 + N_{\text{ex}} = N$, could be controlled by precisely controlling the chemical potential, or the depth of the trapping potential, and, at the same time, accurately counted in each experimental shot by means of a fluorescence imaging technique demonstrated in \cite{RaizenBECstatisticsPRL2005}. A sub-Poissonian level of accuracy for this technique of counting the number of the trapped particles $N$, namely, a relative error less than $1\big/\sqrt{N}$, was achieved in \cite{RaizenBECstatisticsPRL2005} for the traps with $N \sim 10^3$ or less. Potentially, this method allows one to count every single particle in the trap. 

Another interesting technique for local measurements of the sub-Poissonian atom number fluctuations via {\it in situ} absorption images has been demonstrated in \cite{Jacqmin2010} for a 1D Bose gas on an atom chip and revealed the crossover from weak to strong interactions. In particular, for weak interactions, fluctuations go continuously from super- to sub-Poissonian as the density is increased, which is a signature of the transition between the sub-regimes where the two-body correlation function is dominated, respectively, by thermal and quantum contributions. 

There is also a dispersive imaging technique based on the Faraday effect for measuring the number of atoms in a large, ultracold atomic cloud. Its minimally destructive character allows one to take multiple images of the same cloud. The latter enables a sub-atom shot noise measurement precision being itself known {\it in situ} \cite{Kristensen2019,Kristensen2017}.

Remarkably, even the third moment of the atom number fluctuations is accessible for measurements as has been shown in the studies of the density fluctuations in small slices of a very elongated weakly interacting degenerate Bose gas \cite{Armijo2010}. A positive skewness of the atom number distribution in the ideal gas regime and a reduced skewness compatible with zero in the quasicondensate regime have been found. In such a setup, the third moment is a thermodynamic quantity whose measurement constitutes a sensitive test of the equation of state. Of course, the BEC-occupation fluctuations are different from the atom number fluctuations studied in \cite{Armijo2010,Jacqmin2010,Castin,Pit2011,RaizenBECstatisticsPRL2005,AspectDensityFluctPRL2006,SchmittPhotonBECStatisticsPRL2014,StoofBECofLightPRL2014}. Their measurement is a more challenging task. It requires an implementation of special techniques. 

A possible approach is to measure the fluctuations in the superfluid fraction of the BEC-condensed gas. It had been suggested in \cite{Hadzibabic2010} that one could use the optical beams with nonzero angular momenta to simulate uniform rotation of the condensate. The induced change in angular momentum of the BEC-condensed gas can be measured spectroscopically that would result in a direct determination of the superfluid fraction.

An interesting option is to start an experiment on the BEC statistics with a preparation of a Fock state of the trapped particles with some total number of particles $N$ in a strongly interacting regime by a trap reduction based on a combination of weakening and squeezing of the trapping potential \cite{RaizenTrapControl}. Such a nonadiabatic change of the trapping potential leads to a robust preparation of the Fock state for physically realistic traps at finite temperatures both for the flat and smooth trapping potentials. Then, the subsequent nonequilibrium and equilibrium BEC fluctuations could be observed and measured. 

Current experiments are usually based on a resonance absorption imaging of the atomic cloud after some free flight expansion out of the trap triggered by a nonadiabatic change or turning off the trapping potential. Fitting the obtained density profile into a calculated bi-modal distribution of particles in the center and outer clouds is assumed to give the number of particles in the condensate and noncondensate, respectively. 
However, the nonadiabatic change of the trapping potential and subsequent cloud expansion drastically restructure all bare particle states in the empty trap, dressed particle states of the BEC-modified single-particle Schr$\ddot{\text{o}}$dinger equation (\ref{mod_Shr}), and quasiparticle states of the Bogoliubov-de Gennes equations (\ref{GP+BdG}) and considerably modify the effect of the interparticle interaction on the BEC and its fluctuations. 
In particular, the momentum distribution is not preserved in the time-of-flight measurements and is influenced by interactions during the expansion \cite{PitPRA2016,DemlerPRA2017}.
These circumstances make questionable an association of the relative weights of the two spatial modes of the density profile with the numbers of the condensed and noncondensed particles, especially if one assigns particular values to the stochastic variables $N_0$ and $N_{\text{ex}}$ in each experimental shot.
Such an interpretation of the experimental data could be justified to some extend if one addresses the qualitative scaling properties of the variance. A quantitative interpretation requires a further analysis.

\section*{ACKNOWLEDGMENTS} 
The support from the Foundation for the Advancement of Theoretical
Physics and Mathematics “BASIS” and the Russian Science Foundation (grant 18--72--00225) for the work presented in sections I, IV and II, III, V, VI, respectively, is acknowledged. The analysis of the relevant experiments in section VII was performed as part of the State Assignment of the Institute of Applied Physics RAS, project No. 0035-2019-0002. 

%\bibliography{experiments,our-bec-papers,}

{}

\end{document}